\documentclass[a4paper,fleqn,usenatbib]{mnras} 

\usepackage{newtxtext,newtxmath}

\usepackage[T1]{fontenc}
\usepackage{ae,aecompl}

\usepackage{graphicx}	
\usepackage{amsmath}	
\usepackage{booktabs}
\usepackage{color}
\usepackage{pdflscape}
\usepackage{subfig}
\usepackage{multirow}

\newcommand{\ann}[1]{\textit{#1}}	

\newcommand\ExampleHzBreak{1}
\newcommand\ExampleHubbleAge{2}
\newcommand\ExampleMathrmBreak{3}
\newcommand\ExampleOrTextBreak{4}
\newcommand\ExampleRandomBreak{5}
\newcommand\ExampleHoEPM{6}

\title[Towards Machine Learning-Based Meta-Studies]{
    Towards Machine Learning-Based Meta-Studies: Applications to Cosmological Parameters
}

\author[T. Crossland, et al.]{
    Tom Crossland,$^{1,2}$\thanks{E-mail:~\texttt{t.crossland.17@ucl.ac.uk}}
    Pontus Stenetorp,$^{2}$
    Daisuke Kawata,$^{1}$
    Sebastian Riedel,$^{2}$
    \newauthor{}
    Thomas D. Kitching,$^{1}$
    Anurag Deshpande,$^{1}$
    Tom Kimpson,$^{1}$
    Choong Ling Liew-Cain,$^{1}$
    \newauthor{}
    Christian Pedersen,$^{3}$
    Davide Piras,$^{3}$
    and Monu Sharma$^{1}$
    \\
    $^{1}$Mullard Space Science Laboratory, University College London, Holmbury St. Mary, Dorking, Surrey RH5 6NT, United Kingdom\\
    $^{2}$Department of Computer Science, University College London, Gower Street, London WC1E 6BT, United Kingdom\\
    $^{3}$Department of Physics and Astronomy, University College London, Gower Street, London, WC1E 6BT, United Kingdom
}

\date{Accepted XXX. Received YYY; in original form ZZZ}

\pubyear{2020}

\begin{document}
\label{firstpage}
\pagerange{\pageref{firstpage}--\pageref{lastpage}}
\maketitle

\begin{abstract}

We develop a new model for automatic extraction of reported measurement values from the astrophysical literature, utilising modern Natural Language Processing techniques. We use this model to extract measurements present in the abstracts of the approximately 248,000 astrophysics articles from the arXiv repository, yielding a database containing over 231,000 astrophysical numerical measurements.
Furthermore, we present an online interface (\textit{Numerical Atlas}) to allow users to query and explore this database, based on parameter names and symbolic representations, and download the resulting datasets for their own research uses.
To illustrate potential use cases we then collect values for nine different cosmological parameters using this tool. From these results we can clearly observe the historical trends in the reported values of these quantities over the past two decades, and see the impacts of landmark publications on our understanding of cosmology.

\end{abstract}

\begin{keywords}
methods: data analysis -- cosmological parameters -- publications, bibliography -- astronomical data bases: miscellaneous
\end{keywords}




\section{Introduction}

There is currently an unprecedented level of availability of scientific literature and knowledge, made possible by the internet and the open-science spirit of many in the
community. In addition, we are seeing increasing numbers of new publications being added to these repositories at a remarkable rate. Whilst this availability is highly beneficial to the wider community, the sheer number of publications does cause issues for academics wishing to overview literature on particular topics. Due to the technical nature of the domain, keyword search queries and other common content-retrieval algorithms (such as those used by NASA ADS and the arXiv search interface) are often insufficient for identifying useful collections of documents. More than this, if one is searching not just for particular articles, but specific data contained within those articles -- such as numerical measurements, as concerns us here -- the problem is compounded. Not only do we have the task of identifying the relevant papers, but also of reading and cataloguing the data we are interested in. For example, many researchers are regularly interested in meta-studies on the values of specific parameters, where an understanding of the current consensus is required, such as for use in simulations or experimental calculations. When done manually, these endeavours are often time-consuming and can be prone to clerical errors 
and human bias \citep{Kerzendorf2017arXiv170505840K}.

This, therefore, is a task which would benefit from the support of automated approaches, both to free up research time from manual data collection and book-keeping, and also to broaden the horizons of our search -- what with machines not becoming bored after reading the thousandth paper, and not having any unconscious bias towards popular articles. Such a search algorithm could be pre-run over the entire backlog of available literature, allowing for fast search-time queries by users, and then be automatically kept up-to-date as new publications are released.

However, even within the presumably well-structured texts found in scientific literature we still see a vast array of linguistic creativity from the authors of the papers we read.
This presents a multitude of interesting challenges when performing computations on the texts. In particular, more rudimentary algorithms based on heuristics and hand-coded rules often fall short due to the high potential for variation in the patterns one is attempting to capture.

In our previous work \citep{crossland1} we utilised these simpler strategies, in the form of pattern-matching and keyword search, to identify numerical measurements in the astrophysical literature. However, balancing the scope and selectivity of hand-written regular expressions for the large quantity of writing styles seen in the literature is a difficult process, and resulted in large amounts of noise in the results. This in turn required additional hand-tuned filtering steps. These many steps of processing led to gaps in the patterns we were able to capture, and the rule-based nature of the process meant the algorithm was brittle in the face of irregular writing styles.

This problem of variability in free text is well studied in the field of Natural Language Processing, the area of machine learning concerned with the processing of human language, and more modern techniques have been developed in recent years to better capture and understand the complex patterns found in language. Lately this process has been heavily influenced by the development of artificial neural networks, especially in the case of recurrent neural networks, and as such neural techniques for Natural Language Processing have become commonplace in recent years. Models such as BERT/RoBERTa \citep{BERTpaper,RoBERTapaper} and GPT-3 \citep{gpt3paper} are excellent examples of the successes which may be seen from this trend.

Our goal here, then, is to leverage some of these statistical techniques for our problem of numerical measurement extraction from astrophysical literature. This will allow us to overcome some of the shortcomings found in our previous approach, and extend its successes to more complex instances of measurement reporting.
In particular, our earlier attempt was severely limited by the requirement that the parameter names and symbols be pre-specified and atomic (i.e. known to the user and of a singular, rigid form). These requirements simply cannot be met for many real-world parameters, as there is either no single agreed-upon name, perhaps because the entity represents some complex definition (e.g. $\sigma_8$), or no agreed upon symbol. Or worse, the symbol may be over-utilised (e.g. $\beta$). In other cases, the user only has access to an incomplete list of names and symbols, and this limits the recall of their search. In cases such as these a more contextually-aware technique is required, in order to leverage the information available in the text itself.


The final goal of this project is to produce a system which will allow researchers to quickly and easily search the available corpus of literature for instances of measurements of a particular parameter. Our initial investigations towards this goal focused on simple measurement extraction of a single parameter with a well-defined name and symbol (the Hubble constant, $H_0$). In this work we extend this using statistical techniques to a general search for any parameters contained in the text. This means that the ``search'' aspect of utilising the model is moved to a pipeline post-processing step, rather than a user query-time step, which greatly improves efficiency for the user, in addition to providing theoretical advantages for the model structure.

In the following sections we discuss the steps involved in producing these new models, beginning with a brief description of the data we are utilising and the pre-processing pipeline which converts it into an appropriate format (see Section~\ref{sec:data}). For this project we must also create training and evaluation datasets for our task as, to the best of our knowledge, none currently exist.
This will involve the construction of a hand-annotated training dataset created from examples of astrophysical literature, a process which is discussed in Section~\ref{sec:annotationproject}.


Using this training data, we train artificial neural network models to perform the named entity recognition and relation classification tasks for our problem. This will involve identifying spans in the text relating to physical parameters, their mathematical symbols, reported measurements and other numerical data, and so on, and then linking these together such that numerical measurements can be connected to the physical parameters they represent. The architectures and training of these models is discussed in Section~\ref{sec:models}.

\begin{figure*}
	\includegraphics[width=0.9\textwidth]{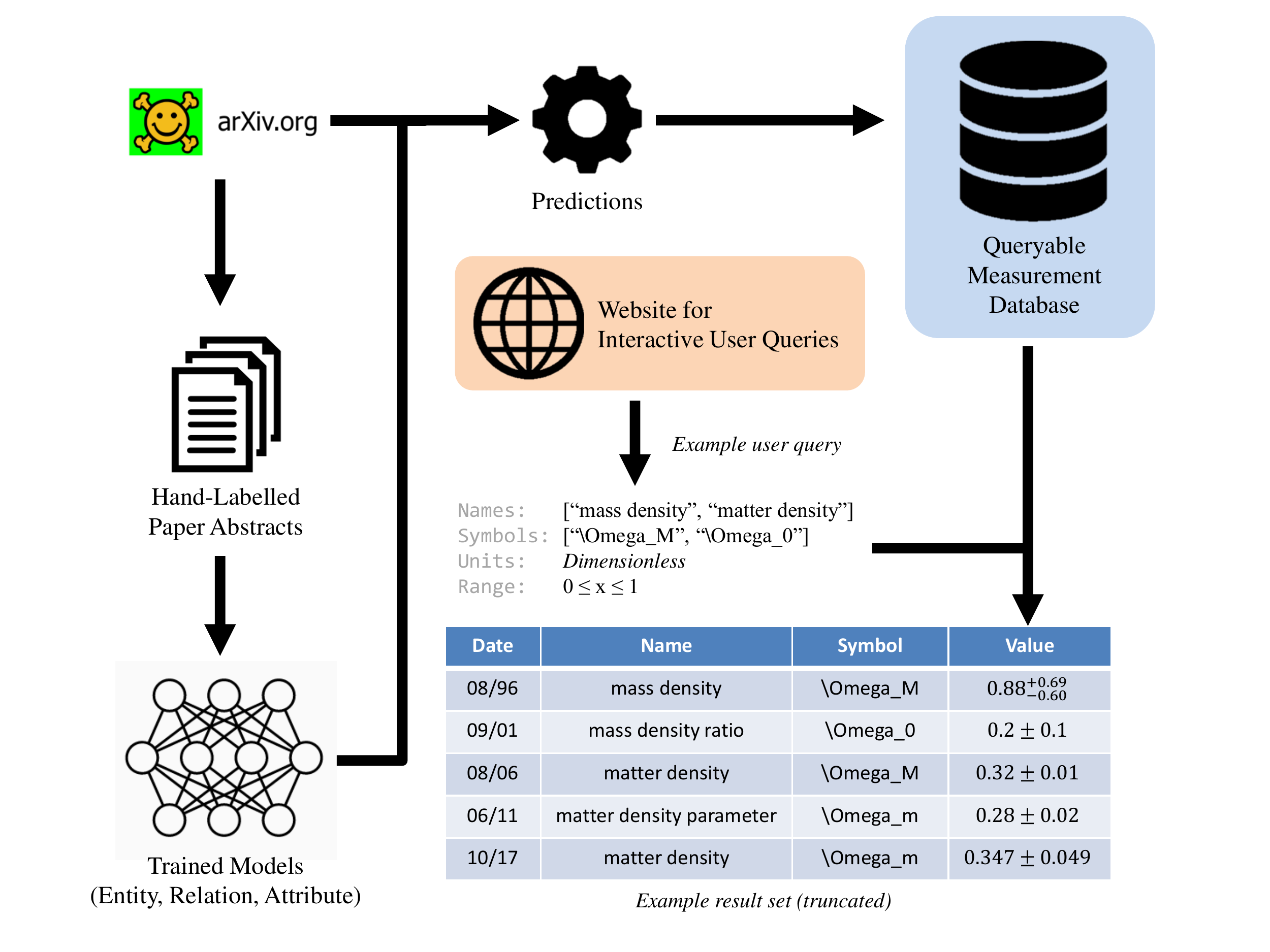}
	\caption{Schematic overview of the project. Using a hand-annotated sample of papers from the arXiv repository, we train a collection of models for measurement extraction, and then perform this data extraction on all the astrophysics paper abstracts in arXiv. These results are then made available via an online interface for interactive user queries.}
	\label{fig:paperfigure}
\end{figure*}

These trained models are applied to the entire arXiv dataset and the outputs used to create a searchable database of numerical measurements which can be easily queried to extract measurements of a given parameter, and other useful information regarding the reporting of such measurements (e.g. confidence limits, constraint values, associated objects). This database will be made available to the community via an online interface, available at \url{http://numericalatlas.cs.ucl.ac.uk}. A schematic diagram of the project outline is shown in Fig.~\ref{fig:paperfigure}.

Finally, Section~\ref{sec:results} focuses on comparing the new statistical approach with the previous rule-based approach from \citet{crossland1}, showing how it performs equally well on the simple tasks that the rule-based model excelled at, and how it surpasses the rule-based approach in more complex situations. We also present a set of example use-cases of the result set, focusing on extracting values of various cosmological parameters, demonstrating the various search parameters which may be employed.
Using these results we discuss some of the trends and features which are observed in the community's understanding of these quantities over the last few decades, and how they may relate to particular events and publications during that time. This will show the utility of the model for scientists wishing to quickly gather numerical information relating to a measureable physical quantity for various kinds of analyses.


\section{Data}\label{sec:data}
    
The dataset for this project is taken from astrophysics publications from the arXiv, an open-source repository for scientific literature, maintained by Cornell Tech\footnote{\url{https://www.tech.cornell.edu/}}. Publications on the arXiv may be stored in a variety of formats, with the most common being \LaTeX{} source files (91\% of all submitted articles). As such, we have chosen to utilise the structured nature of the \LaTeX{} files to allow us to process the documents into well-formatted text appropriate for machine learning tasks.

In order to process these source files into a more usable format, we have utilised the pre-processing pipeline described in \citet{crossland1}.
Article source files are processed using the LaTeXML program, created by the National Institute of Standards and Technology\footnote{\url{https://dlmf.nist.gov/LaTeXML/}}, into a single XML document, which improves the usability of the data for computational purposes. The text is then tokenised and sentence split -- with a purpose built tokeniser for \LaTeX{} \texttt{math} environments. Using this corpus we can easily create textual data samples to a variety of specifications for our machine learning models, based on the content of section headings (e.g. ``Results'' or ``Conclusions''), document components (e.g. \texttt{abstract}), and so on.

Our current dataset consists of all arXiv papers published up until September 2020, corresponding to 1.6 million articles. Of these, approximately 265,000 have the astrophysics tag (``astro-ph'), and our pipeline can successfully extract over 248,000 formatted articles from this set (corresponding to a success rate of 94\%). Failure cases are generally found in older articles, often due to the source files being written in \TeX{} rather than \LaTeX{}. This coverage of the available articles is considered to be sufficient for the purposes of this task, and it shall be assumed in this work that the 94\% of processed articles are statistically similar to the remaining data, in terms of their content and linguistic style.

We have also utilised the dataset compiled by \citet{Croft+Dailey2011arXiv1112.3108C}, which comprises 638 values of 8 cosmological parameters from 468 papers. These papers are used as a curated set of example literature for our task, both for analysis and as a component of the annotation effort described in Section~\ref{sec:annotationproject}.

We have made one further assumption in the use of this data: that any paper whose goal is to report some numerical measurement as a finding of the publication will report said measurement in the paper abstract. This is not always the case, especially for publications concerning the determination of numerical quantities for a set of objects (stellar parameters for some large sample of stars, for instance). However, based on our investigations of the \citet{Croft+Dailey2011arXiv1112.3108C} dataset, we find this to be a reasonable working assumption and use it for the majority of this work. Specifically, this makes the creation of a manually annotated training corpus a more tractable proposition. It is, however, noted that there are distinct linguistic differences between article abstracts and main bodies, and generalising the models trained on this data to entire papers will be the subject of future work.

\section{Annotation of Astrophysics Abstracts}\label{sec:annotationproject}


For our machine learning tasks we require data to train and evaluate our models -- examples which show the mapping between input data and desired output. Therefore the next step in our data processing is to produce hand-annotated samples which demonstrate the information we wish our models to extract (annotated article abstracts in our case).

In Natural Language Processing there are many kinds of annotation which may be produced; here we are interested in Entity, Relation, and Attribute annotations.

An \textbf{Entity annotation} is one where we select a span of text from our document
and assign some label to that span. For example, in the sentence, ``\ldots{}for the Hubble constant at the present epoch\ldots{}'', we could select the span ``Hubble constant'' and assign it the label \ann{ParameterName}.

A \textbf{Relation annotation} is where we have two Entity annotations and we declare the existence of some semantic relationship between them. For example, in the sentence, ``Using the \textbf{Hubble constant}, $\mathbf{H_0}$, under the assumption\ldots{}'', we could create a Relation between the Entities ``Hubble constant'' (\ann{ParameterName}) and ``$H_0$'' (\ann{ParameterSymbol}), and assign it the label \ann{Name} (the labels used in this project are discussed below). Relation annotations may be constrained by the Entity types which they may connect. For example, a \ann{Name} Relation may only exist between a \ann{ParameterName} Entity and a \ann{ParameterSymbol} Entity (generally, these constraints are not symmetric, meaning that most Relations are directional).

Finally, an \textbf{Attribute annotation} is one which modifies an Entity, by assigning another label to it. For example, in the sentence, ``Using a value of \textbf{0.3} from the literature\ldots{}'', we could assign a \ann{LiteratureValue} Attribute to the Entity ``0.3'' (\ann{MeasuredValue}). As for Relations, Attributes may be constrained by the type of Entity they can be assigned to. For example, a \ann{LiteratureValue} Attribute can only be placed on a \ann{MeasuredValue} or \ann{Constraint} Entity.

Now that we have our annotation types, we create a schema which describes the Entity, Relation and Attribute labels we have available for our annotation project, and the constraints which exist for them. We are interested in measurement extraction from astrophysical literature, and so require labels which reflect that domain: Entity labels for measurements, parameters, objects and definitions are all appropriate. Likewise, for Relations, we must be able to define which names and symbols relate to which measurements, and which parameters are properties of which objects, and so on. A complete list of the annotations used in this project may be found in Tables \ref{tab:entities}, \ref{tab:relations} and \ref{tab:attributes}, along with any constraints which exist on them. Detailed descriptions of each may be found in Appendix~\ref{app:anndescriptions}.
This schema is not intended to represent an exhaustive list of the various semantic entities which may be relevant to this problem or domain. A compromise has been struck between completeness and practicality, as we will be requiring human annotators to implement this schema when annotating training data (as a very detailed schema is impractical for annotators, if there are too many annotation types and combinations to remember). As such, we have chosen to focus on the most important Entities and Relations for our task, favouring broader definitions over an increased number of labels in certain cases (e.g. \ann{ObjectName} labels, where we could easily have multiple labels for different kinds of physical entities).

\begin{table}
\centering
\caption{Entity annotation types in the annotation schema. Detailed descriptions of these may be found in Appendix~\ref{app:anndescriptions}.}
\label{tab:entities}
\begin{tabular}{ll}
\hline
\textbf{Name}        \\
\hline
MeasuredValue        \\
Constraint           \\
ParameterName        \\
ParameterSymbol      \\
ObjectName           \\
ConfidenceLimit      \\
\hline
\end{tabular}
\end{table}

\begin{table}
\centering
\caption{Relation annotation types in the annotation schema, showing any constraints on the start and end Entity types for the listed Relations. Note that [Measurement] refers to either a \ann{MeasuredValue} or \ann{Constraint} annotation, and [Parameter] refers to either a \ann{ParameterName} or \ann{ParameterSymbol} Entity annotation. See Appendix~\ref{app:anndescriptions} for detailed descriptions.}
\label{tab:relations}
\begin{tabular}{llll}
\hline
\textbf{Name} & \textbf{Start} & \textbf{End}                \\
\hline
Measurement   & [Parameter]    & [Measurement]               \\
Name          & ParameterName  & ParameterSymbol             \\
Confidence    & [Measurement]  & ConfidenceLimit             \\
Property      & ObjectName     & [Parameter] | [Measurement] \\
Equivalence   & ObjectName     & ObjectName                  \\
Contains      & ObjectName     & ObjectName                  \\
\hline
\end{tabular}
\end{table}

\begin{table}
\centering
\caption{Attribute annotation types in the annotation schema, showing any constraints on the subject Entity type. Note that [Measurement] refers to either a \ann{MeasuredValue} or \ann{Constraint} Entity annotation. See Appendix~\ref{app:anndescriptions} for detailed descriptions.}
\label{tab:attributes}
\begin{tabular}{lll}
\hline
\textbf{Name}  & \textbf{Entity} \\
\hline
Incorrect      & [Measurement]   \\
AcceptedValue  & [Measurement]   \\
FromLiterature & [Measurement]   \\
UpperBound     & Constraint      \\
LowerBound     & Constraint      \\
\hline
\end{tabular}
\end{table}

\subsection{Annotation Process}

Using this schema and a team of 7 astrophysics PhD student annotators we have annotated 600 article abstracts, with each abstract being annotated by 3 annotators. For this process we utilised the brat rapid annotation tool \citep{bratpaper}. The resulting set of annotations have then been combined such that each abstract has a single, consensus annotation set, and it is this consensus data which will be utilised as training data by our machine learning models. The steps taken in this process are detailed below.

Firstly, we select a set of papers to be annotated from the available corpus. As a starting point we choose the 305 papers contained in the \citet{Croft+Dailey2011arXiv1112.3108C} dataset (the subset that successfully pass through our preprocessing pipeline, as discussed in Section~\ref{sec:data}). These serve as examples of the papers reporting measurements of cosmological parameters that we wish to identify in our test cases, as in Section~\ref{sec:results}. To round out this selection of papers, we score the available papers from the arXiv dataset according to an estimate of the number of measurements used in the paper abstracts \citep[for this estimate we use a regular expression to identify candidate measurement strings in the text, as described in Section~4.2 in][]{crossland1}. We then filter these measurements to remove noise, notably by requiring that the measurement patterns contain uncertainties.
Due to the prevalence of dimensionless quantities in cosmology, we also reject papers which only contain measurements with concrete units, such that the distribution of these papers will be closer to that of the \citet{Croft+Dailey2011arXiv1112.3108C} dataset. We then randomly sample papers with a non-zero estimated number of measurements in their abstracts
to produce a final set of papers for annotation.

It should be noted, therefore, that this set of papers is heavily biased towards cosmological measurements, and this will have an impact on the efficacy of the model in identifying measurements in other areas. However, we should also note that the randomly sampled papers are not constrained by arXiv subject tag (beyond simply the astrophysics tag, ``\texttt{astro-ph}''), and so are selected from a range of subject areas within astrophysics. This bias was chosen due to the target test case for this work being cosmological parameters (see Section~\ref{sec:resultscosmoconstants}).

For the annotation project itself we recruited 7 astrophysics PhD students and presented them with a set of example annotated documents based on the schema outlined above. The selected papers were then released in batches of 100, evenly divided between the \citet{Croft+Dailey2011arXiv1112.3108C} and randomly sampled papers, over the course of several months. The annotators were paid for their time at their standard rate, allowing for an average of 5 minutes per abstract. The papers were allocated such that each was annotated by three separate annotators.

Each round of annotations was conducted in two stages: first, the annotators were asked to work independently on their sample, and secondly, once these initial annotations were complete, they were made available to all annotators, who were then asked to compare their annotations with the others and bring the annotations for each paper into better alignment. However, it should be noted that it was not a requirement that the annotators ensure their annotations be in perfect agreement, meaning that the final dataset still contains some discrepancies between individual annotation attempts. This approach was used to ensure that the final dataset benefited from the different perspectives of the annotators, whilst also ensuring that the final result represented the considered opinion of multiple domain experts.
These repeated annotations were then consolidated into single annotation sets, representing the consensus of the annotators.

Our final dataset contains 572 paper abstracts (after accounting for papers which were unsuitable, contained no useful annotations, or found to be incorrectly formatted), with 17,446 annotations (10,352 Entities, 6,447 Relations, 647 Attributes) after post-processing (see Appendix~\ref{app:consensusalgorithm}). This is comparable to other existing datasets, such as the CoNLL-2002 Shared Task \citep[a Named Entity Recognition dataset, with 6,655 and 6,299 Entities for the Spanish and Dutch categories, repsectively, presented by][]{conll2002} and SemEval-2010 Task 8 \citep[a Relation classification dataset with 10,717 Relations, presented by][]{semeval2010task8}.

\subsection{Annotation: Caveats}\label{sec:annotationcaveats}


There were a few issues encountered during the annotation process which should be noted: Firstly, the \ann{ParameterName} annotation causes some issues with agreement between annotators. This is to be expected, as the exact span of a parameter's name can be difficult to define exactly. Some examples of this would be, ``mean baryon density of the Universe'', ``total mass of three massive neutrinos'', or, ``mass-weighted Galactic disk scale length'' (all examples taken from our annotated documents). In these instances, there is a more compact span which could approximate the `name' in question (``baryon density'', ``total mass'', ``scale length''), but does not accurately capture the full intended context. We can, of course, generally extend this reasoning in both directions arbitrarily far -- right down to single words, and up to full sentences (or even paragraphs) of explanation -- but this is often impractical. Deciding on the exact compromise is difficult, and this leads to different annotators selecting slightly different spans for many instances of \ann{ParameterName} annotations. The alignment segment of our annotation strategy alleviates this disagreement somewhat, but it serves to show that this Entity has a lot of linguistic ambiguity. Indeed, we shall see in Section~\ref{sec:models} that our models struggle to achieve higher scores when recognising these labels -- a combination of these disagreements between annotators carrying over to the dataset, and the inherent linguistic ambiguity in the boundaries of these phrases.

We also see issues with the \ann{ObjectName} annotations. In some instances this is closely related to the problems with \ann{ParameterName} boundaries. For example, the phrase, ``mass-weighted Galactic disk scale length'', could be annotated as a single \ann{ParameterName}, or as the \ann{ParameterName} ``scale length'' which is in turn a \ann{Property} (Relation) of the \ann{ObjectName} ``Galactic disk''. If the phrase had been written, ``Milky Way disk scale length'', this breakdown into \ann{ObjectName} and \ann{ParameterName} would perhaps be more appealing, but the use of an adjective (``Galactic'') coupled with a self-contained phrase (``disk scale length'') may give the annotator pause. Context is also important in many of these situations, as reference to a simulated object rather than an observed one may bias the annotator away from using an \ann{ObjectName} label, and so on.

It should be noted, however, that the combination of annotator discussion and our consensus algorithm (see Appendix~\ref{app:consensusalgorithm}) go a long way to alleviating the observed disagreements. They are discussed here to illustrate the problem cases presented by the data, and the problems we will encounter during model training.


\section{Models}\label{sec:models}

\subsection{Tasks}

We have chosen to formulate the overall task of finding measurements in free text as two sub-tasks, which are both well-documented in the Natural Language Processing domain: Named Entity Recognition \citep{nercSurveyPaper} and Relation Classification \citep{2017arXiv171205191P}. In contrast to our approach in \citet{crossland1} we approach these tasks using artificial neural network techniques, as has become standard practice in Natural Language Processing in recent years, rather than the heuristic approach taken before. This can give the models more flexibility and scope, allowing for a broader investigation of the data available in the literature (however, as we shall see, this cannot always overcome the inherent difficulty of the task, as seen with our neural Relation models).
Additionally, we have a simpler classification task for predicting Attributes. Other than the inclusion of a recurrent neural network to deal with the variable length sequences involved, this will be formulated as a traditional classification problem.

\subsubsection{Named Entity Recognition} \label{sec:namedentityrec}

In Named Entity Recognition tasks we consider the text as a series of individual tokens, which may be words, numbers, punctuation marks, or other self-contained collections of characters (without whitespace). The task is then to find subsequences of these tokens which correspond to named entities. In general tasks, this may be place or person names (often consisting of multiple tokens, for example, ``Hubble Space Telescope''), or any other sequence of tokens which together refer to some single entity. For example, the Entity ``effective temperature'' (with label \ann{ParameterName}) is comprised of the tokens ``effective'' and ``temperature'', whereas the Entity ``H \_ \{ 0 \}'' (\ann{ParameterSymbol}) consists of the tokens ``H'', ``\_'', ``\{'', ``0'', and ``\}''. Named Entity Recognition is distinct from the task of assigning labels to individual tokens, such as labelling words as ``\ann{verb}'', ``\ann{noun}'', ``\ann{adjective}'', etc. in a sentence, a task generally referred to as Part-of-Speech tagging. The list of named entities we are considering in this work are the same as those found in Table~\ref{tab:entities}.

A common practice in Named Entity Recognition tasks is to classify tokens according to the Beginning-Inside-Outside (BIO) format \citep{ramshaw-marcus-1995-text}, where each token is designated as either a ``beginning'' token (corresponding to a particular label, for example, ``<B-ParameterName>''), an ``inside'' token (again corresponding to a particular label, e.g. ``<I-ParameterName>''), or an outisde token (not belonging to any label, ``<O>''). An example sentence showing this labelling is given in Table~\ref{tab:examplesentence}.

\begin{table}
\centering
\caption{Example BIO-labelled tokenenised sentence.}
\label{tab:examplesentence}
\begin{tabular}{ll}
\hline
\textbf{Token}  & \textbf{Label} \\
\hline
We & Outside \\
find & Outside \\
the & Outside \\
Hubble & B-ParameterName \\
constant & I-ParameterName \\
to & Outside \\
be & Outside \\
70 & B-MeasuredValue \\
km & I-MeasuredValue \\
/ & I-MeasuredValue \\
s & I-MeasuredValue \\
/ & I-MeasuredValue \\
Mpc & I-MeasuredValue \\
. & Outside \\
\hline
\end{tabular}
\end{table}

Hence, for a set of $N$ Entity names, we have a possible $2N+1$ BIO labels (``begin'' and ``inside'' for each Entity name, and one ``outside'' label). This, therefore, is the number of output classes for our machine learning models.

The BIO format has some drawbacks in the general case, notably that it cannot express nested or overlapping annotations, but as we have specified that our Entity annotations will be non-overlapping we will not encounter this problem here.

\subsubsection{Relation Classification}

Relation Classification is the subject of much active research in the field of Natural Language Processing. Many of the recent works in this field have involved datasets comprised of single-sentence samples, where each sample either contains one of a set of possible Relations, or no Relation at all \citep[e.g.][]{semeval2010task8}. However, we cannot easily break our data down into these atomic relational chunks, as we have many sentence which contain multiple Relations,
and many long-distance Relations (where the Entities are not contained in the same sentence, and may even be several sentences apart in the text). Therefore here we are considering the task of relation classification between labelled Entities in free text. The exact formulation of this problem is treated differently for the models described below, and so will be discussed in following sections.



\subsection{Featurization}

All of the models we use in this project require a mechanism for converting tokens into a numerical vector representation, often referred to as an embedding. These embeddings may then be used in mathematical operations, such as the matrix operations which underlie all neural network layers. There are many algorithms and models currently in use for this purpose, such as Word2Vec \citep{word2vecRelease}, GloVe \citep{GloVepaper}, or BERT/RoBERTA \citep{BERTpaper,RoBERTapaper}. We have chosen to use Word2Vec, as it is a class of models which are well documented, and can be retrained locally if a large corpus is available (such as our arXiv dataset). Word2Vec operates by creating an embedding space (vector space), where each token in the vocabulary is assigned a separate vector representation. The Word2Vec model is then trained such that ``similar'' words have similar embeddings -- i.e. appear close to each other in the embedding space.
Word2Vec is a very powerful technique, as the resulting models produce embedding spaces where tokens are clustered semantically and in a structured manner, such that both direction and position have semantic meaning \citep{mikolov-etal-2013-linguistic}.

One downside of Word2Vec is that tokens are defined solely by their character strings. This means that, for example, the words ``play'' as in ``theatrical production'' and ``play'' as in ``play a sport'' only have one embedding, despite having separate meanings -- and the Word2Vec algorithm must encode both possible meanings into a single representation. More recent approaches in Natural Language Processing have utilised contextual word embeddings (e.g. BERT), where the surrounding tokens are taken into account when constructing an individual token's embedding, but these come with a significant runtime and memory cost.

For this project we trained a set of Word2Vec embeddings on the entire arXiv astrophysics corpus (see Section~\ref{sec:data}), and these embeddings will be used for all of the models discussed below.
For efficiency reasons, these embeddings are fixed at training time. However, this can impose limitations on any model using the embeddings (especially shallower networks) and so each model also performs an initial projection of the vectors. This is done with a simple matrix multiplication with a square matrix, which is itself a trainable part of the model. This increases the model capacity with regards to the fixed input embeddings, whilst maintaining the efficiency of pre-trained embeddings.

Whilst the Word2Vec token embeddings provide an excellent basis, they do fall short under certain circumstances. A notable instance of this is in the case of rare tokens -- i.e. specific sequences of characters which occur infrequently. As the Word2Vec algorithm requires a minimum number of occurrences before a token is included in the vocabulary, rare tokens are often referred to as ``out-of-vocabulary'', and are replaced with a default embedding. As our Word2Vec model was trained specifically on astrophysical literature, we are less concerned with out-of-vocabulary technical language, but instead are concerned with numerical strings.

To a human reader, the difference between the strings ``0.70'' and ``0.71'' is minor, as we interpret the value in its numerical sense. However, the Word2Vec algorithm is not designed to leverage the numerical nature of the strings, as they are considered only as a string of characters. Whilst Word2Vec does indeed organise numerical strings in a structured manner, due to their usages in text, this is only sufficient for common numerical strings (``1'', ``15'', `100'', and so on). In our scientific context, important numerical values (especially measurement values) are likely to be rare character sequences. As such, Word2Vec may encounter issues dealing with these tokens \citep{2021arXiv210313136T}.

In order to alleviate this problem, and generally increase the capacity of our Entity models, we have also created versions of the above models which utilise boosted token embeddings. For these models, the embedding for each token is constructed by concatenating the Word2Vec embedding with the output of a trainable character-level neural network encoder \citep[akin to][]{seo2018bidirectional}.

This encoder is a simple single-layer bidirectional long short-term memory (LSTM) network, which is passed over a word matrix, created by concatenating trainable character embeddings. Hence, for a word $W$ of length $w$, with character embeddings of dimensionality $c$, each word may be represented by a $w \times c$ matrix. The hidden state of the Bi-LSTM at the final timestep is used as a fixed-length character-based word embedding for $W$.

Therefore, for these boosted models, each (projected) Word2Vec word embedding is concatenated with the character-based word embedding before being supplied to the model. Training signal is allowed to backpropagate into the character encoder during training, allowing the model to learn to fill in the information gaps in the Word2Vec embeddings, whilst still having the power of the Word2Vec algorithm to fall back on.







\subsection{Data Usage}

When training the following models we use a holdout dataset comprising of all the annotations contained in a subset of the article abstracts from our annotated dataset. This means that the training data for the Entity and Relation models come from the same set of papers, which are distinct from the set of papers used as a holdout testing set. This is done to prevent contamination of the validation results.

\subsection{Entity Models}

To begin, we examine the Named Entity Recognition models we have created: a feed-forward neural network, and a recurrent neural network using LSTM layers. Here we are experimenting with multiple model architectures to give us insights into the complexity of the problem, and aid in interpreting model performance (as the different architectures emphasise different kinds of information from the text).

It should be noted that, due to the relative sparsity of Entities in the texts, for all the models here we shall be combining \ann{MeasuredValue} and \ann{Constraint} Entities for the purposes of token prediction. This improves the model performance on the Named Entity Recognition task, and the \ann{Constraint} annotations can be recovered by using the Attribute model to predict the presence of constraints (i.e. any \ann{MeasuredValue} Entity for which \ann{LowerBound} or \ann{UpperBound} Attribute is predicted can be assumed to be a \ann{Constraint} annotation).



\subsubsection{Feed-Forward Model}

Our first model uses a multi-layer perceptron (MLP) neural network to predict BIO labels for each token in a document. This architecture is a natural baseline for experiments with neural models. We step through the document token by token (starting from the beginning) considering each token's word embedding, concatenated with the embeddings of the tokens in a fixed-width window (forwards and backwards) around the current token, to predict the label for that token. A fixed-length history of previous output predictions is maintained (whose length is equal to the window width) which is also used as input in each prediction step.

\begin{figure}
    \centering
	\includegraphics[width=\columnwidth]{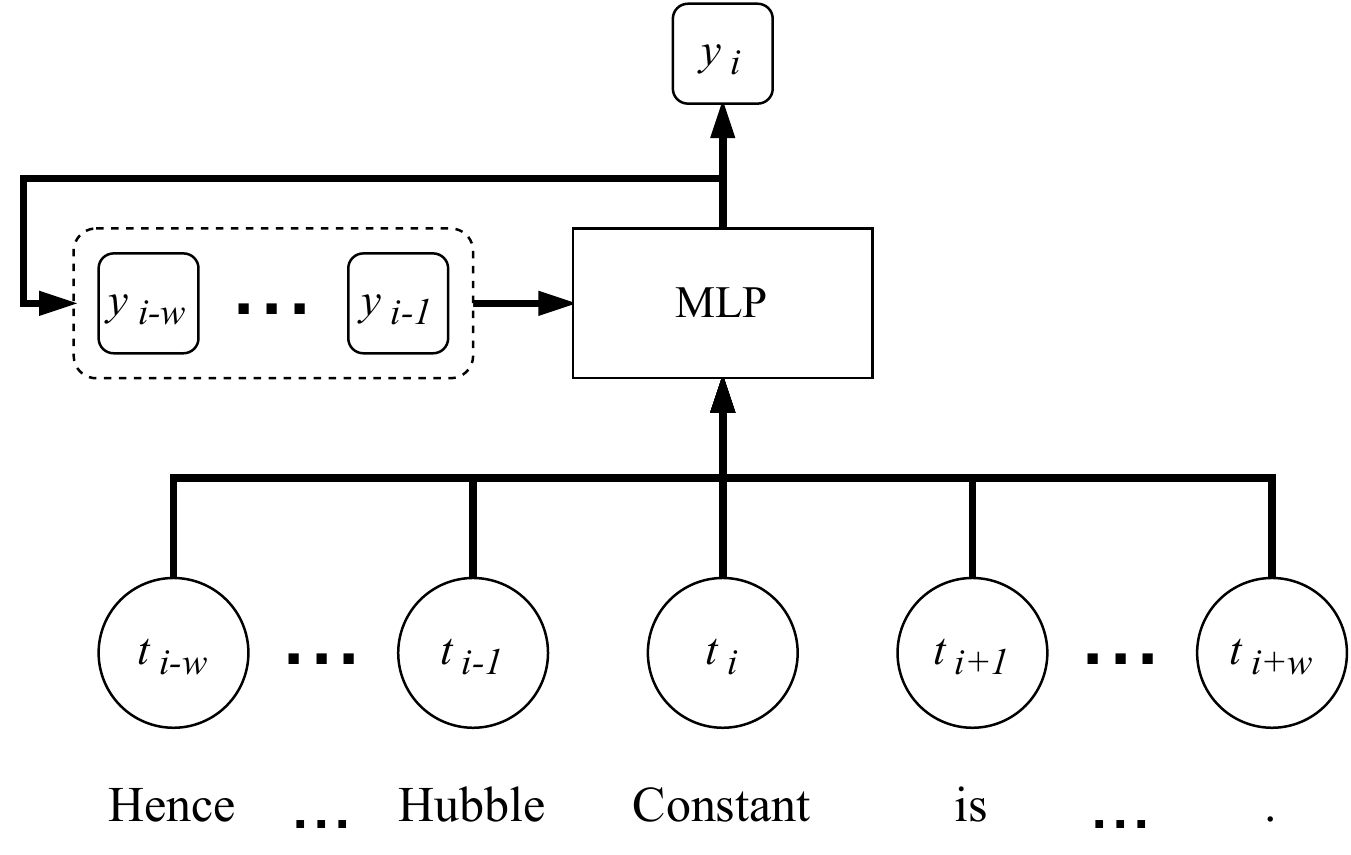}
	\caption{A schematic diagram of the feed-forward Entity model, where $t_n$ indicates the $n^{th}$ token embedding, $y_n$ indicates the model output for the $n^{th}$ token, and $w$ is the window width.}\label{fig:mlpentmodel}
	\label{fig:denseentmodel}
\end{figure}

A schematic diagram of this model is shown in Fig.~\ref{fig:mlpentmodel}. For a model with a window width, $w$, we concatenate the token embeddings of the $2w+1$ tokens in the current window ($w$ to either side, plus the current token) along with the previous $w$ outputs (each a $2N+1$ vector representing the BIO Entity labels, normalised using the softmax function) to produce our input. The prediction history is initialised using a trainable vector parameter, and zero-padding is used to account for the window width (as we begin at first token, not the $w^{\textrm{th}}$ token). The input is then passed through a MLP network, using ReLU activations \citep{reluPaper}, to produce our token label prediction. The exact number of layers and neurons in the MLP network is determined via grid search, with the results for the best performing model shown below.

It should be noted that this model is not a recurrent neural network, despite utilising the outputs from previous tokens, as the training signal is not allowed to backpropagate between token steps. However, the use of the output label ``memory'' was found to greatly improve the model performance.

\subsubsection{LSTM Model}

Our second model uses an LSTM \citep{LSTMpaper} architecture followed by a dense output layer. A schematic diagram of the architecture is shown in Fig.~\ref{fig:lstmentmodel}.
We chose a bidirectional \citep{10.1109/78.650093} LSTM model in this case,
as information will need to propagate in both directions through the text (for example, it is important if a number is followed by a ``$\pm$'' sign, as well as whether it is preceeded by an equals sign). The exact number of layers and cells in the LSTM network is determined by grid search, with the best model performance given below.

For this model, the bidirectional LSTM units are passed along the document, and the sequential output from the LSTM (corresponding to each token) is then sent through a dense output layer, giving the desired $2N+1$ output nodes for each timestep.

The LSTM units should allow the model to capture longer distance dependencies between words and phrases, as it is not limited by a fixed-length window, creating smoother predictions across tokens -- as models without any contextual awareness tend to produce very fractured prediction sequences, where many Entities are incomplete and split due to individual missing tokens.

\begin{figure}
    \centering
	\includegraphics[width=0.8\columnwidth]{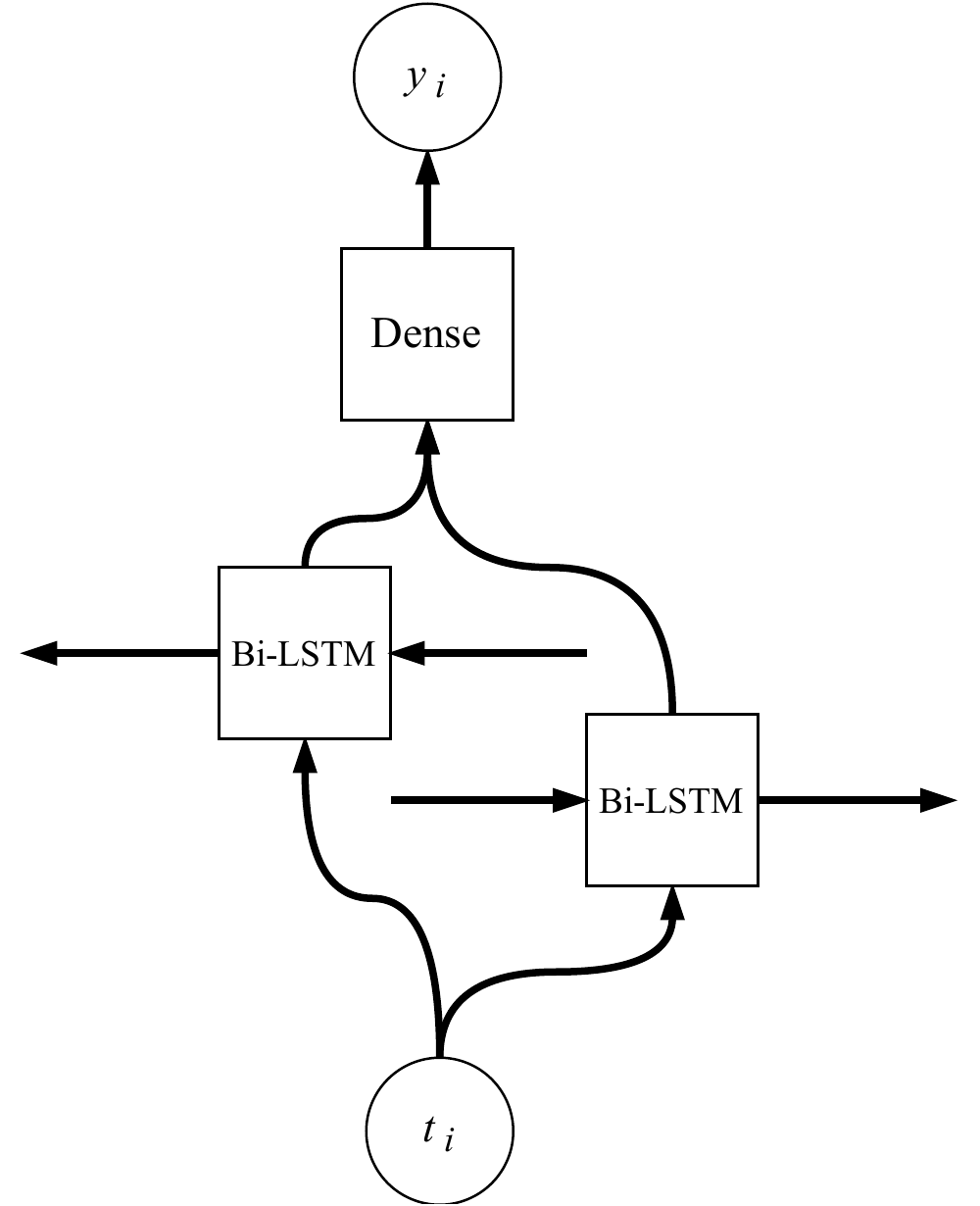}
	\caption{A schematic diagram of the LSTM Entity model. Here, the Bi-LSTM node is the same node in both cases, evaluated forwards and backwards across the text token sequence. Here $t_n$ and $y_n$ indicate the token embedding and model output for the $n^{th}$ token, respectively.}\label{fig:lstmentmodel}
	\label{fig:lstmentmodel}
\end{figure}

\subsubsection{Entity Model Results}

A grid search was performed over the hyper-parameters for both models, with model performance judged using the F1 score and strict Entity overlap (the proportion of Entities which are exactly predicted by the model, i.e. with no missing or additional tokens) on the holdout test dataset. The highest performing models for both proposed architectures were then selected, and their performance statistics are shown in Tables \ref{tab:entsummaryresults} and \ref{tab:entclassresults}.

We see that the two models show comparable performance on this task, with the LSTM model proving slightly more effective overall. This suggests that the linguistic markers required to determine the nature of a token are predominantly local, as the LSTM's capacity to examine longer distance dependencies does not have a particularly large impact on model performance. Indeed, the top performing feed-forward model uses a window-length of only 3 tokens. However, on balance, we have chosen the LSTM model to be used for our final processing steps.

We also note that both models struggle particularly with \ann{ParameterName} and \ann{ObjectName} tokens. In the case of \ann{ObjectName} tokens, this may be explained by the relative sparsity of these Entities in the training data. The difficulties with \ann{ParameterName} labels, however, is suspected to be due to the intrinsic difficulty of separating these tokens from general physical discussion, as well as the ambiguity in the start and end points of these Entities, as shown by the disagreements experienced between annotators during the creation of the training data (see Section~\ref{sec:annotationcaveats}). As seen in Table~\ref{tab:entclassresults}, the models struggle more with recall for these \ann{ParameterName} labels (although the precision is also noticeably lower than for other classes), suggesting that the model predictions represent a more conservative view of what constitutes a parameter name. As such, usage of the outputs for search purposes should emphasise parameter symbols to have the best results.

\begin{table}
\caption{Summary metrics on the test set, for the best performing Entity models from the grid search.}\label{tab:entsummaryresults}
\centering
\begin{tabular}{lccccccc}
\hline
Model Type & Precision & Recall & F1 & Strict Entity Overlap \\ \hline
Feed-forward & .933 & .933 & .933 & .546 \\
LSTM & .934 & .934 & \textbf{.934} & \textbf{.584} \\
\hline
\end{tabular}
\end{table}

\begin{table}
\caption{Per-label performance metrics on the test set, for the best performing Entity models from the grid search. Here MLP refers to the feed-forward model.}\label{tab:entclassresults}
\centering
\begin{tabular}{clcccr}
\hline
\textbf{Model} & \textbf{Relation Type} & \textbf{Prec.} & \textbf{Recall} & \textbf{F1} & \textbf{Support} \\
\hline
\multirow{11}{*}{MLP} & B-ConfidenceLimit & .78 & .77 & .78 & 52 \\
 & I-ConfidenceLimit & .81 & .76 & .79 & 55 \\
 & B-MeasuredValue & .83 & .87 & \textbf{.85} & 536 \\
 & I-MeasuredValue & .94 & .90 & \textbf{.91} & 3,077 \\
 & B-ObjectName & .74 & .61 & .67 & 214 \\
 & I-ObjectName & .80 & .62 & \textbf{.70} & 172 \\
 & B-ParameterName & .62 & .43 & .51 & 471 \\
 & I-ParameterName & .64 & .44 & .53 & 762 \\
 & B-ParameterSymbol & .87 & .85 & .86 & 687 \\
 & I-ParameterSymbol & .84 & .94 & .89 & 2,501 \\
 & Outside & .96 & .96 & .96 & 29,741 \\
 \hline
\multirow{11}{*}{LSTM} & B-ConfidenceLimit & .79 & .88 & \textbf{.84} & 52 \\
 & I-ConfidenceLimit & .78 & .84 & \textbf{.81} & 55 \\
 & B-MeasuredValue & .82 & .86 & .84 & 536 \\
 & I-MeasuredValue & .92 & .90 & .91 & 3,077 \\
 & B-ObjectName & .72 & .68 & \textbf{.70} & 214 \\
 & I-ObjectName & .68 & .61 & .64 & 172 \\
 & B-ParameterName & .60 & .49 & \textbf{.54} & 471 \\
 & I-ParameterName & .61 & .60 & \textbf{.61} & 762 \\
 & B-ParameterSymbol & .84 & .89 & \textbf{.86} & 687 \\
 & I-ParameterSymbol & .87 & .95 & \textbf{.90} & 2,501 \\
 & Outside & .96 & .96 & .96 & 29,741 \\
 \hline
\end{tabular}
\end{table}

\subsection{Relation Models}

For our Relation Classification task we have created two models: a neural network model which considers the two Entities and the span which exists between them (along with a windowed region outside) to classify the Relation that may exist between them; and a rule-based model, which does not use any neural network techniques, but relies on hand-coded heuristics. This rule-based approach was not possible previously, as we did not have access to the token-level predictions from the Entity model which are the basis for the heuristics. We are experimenting with both approaches in order to best explore the possible benefits of the neural model against the interpretability of the rule-based model, to better contextualise model performance.

\subsubsection{Neural Relation Model}

For this model we consider each potential pair of Entities separately, also considering both possible directions of the Relation (as most Relations are directional, and so $A \rightarrow B \neq B \rightarrow A$ in most cases). For every pairing of Entities, $E_m$ and $E_n$ (where $m < n$) we have certain obvious information available: the tokens comprising each Entity span, the labels of these Entities, the tokens of the span between the two Entities, and the labels of the tokens in that span. Additionally, we will use an outer window around the two Entities (i.e. a fixed-length span of tokens which lie outside the Entities and their connecting span) as input into the model, along with any Entity labels which may apply. With this, we have five spans of tokens \citep[akin to][]{hashimoto-etal-2015-relationclassif}. To account for possible directionality of the Relation, we also include a bit indicating whether the Relation runs from the earlier to the later Entity, or vice versa, and evaluate the available spans twice, with differing values for this ``direction bit''. The output of the model is an $N+1$ dimensional vector, where $N$ is the number of Relation labels we are considering (plus one for a ``none'' label).

We now encounter a problem in that there is no predetermined fixed length for the Entity and connecting spans -- they can have any number of tokens (even zero, in the case of the connecting span). As such, we require a way of converting these variable length token matrices (produced by concatenating the token embeddings) into fixed-length representations. We have chosen to use an LSTM for this purpose, where the fixed-length representation is the hidden state of the LSTM at the final timestep (token). Other approaches were experimented with, notably the strategy of taking minimum, maximum, and mean values along the time axis (i.e. the document length) to produce fixed length summary vectors. However, this approach suffers with long distance dependencies, and was out-performed by the LSTM summarisation.

\begin{figure}
    \centering
	\includegraphics[width=\columnwidth]{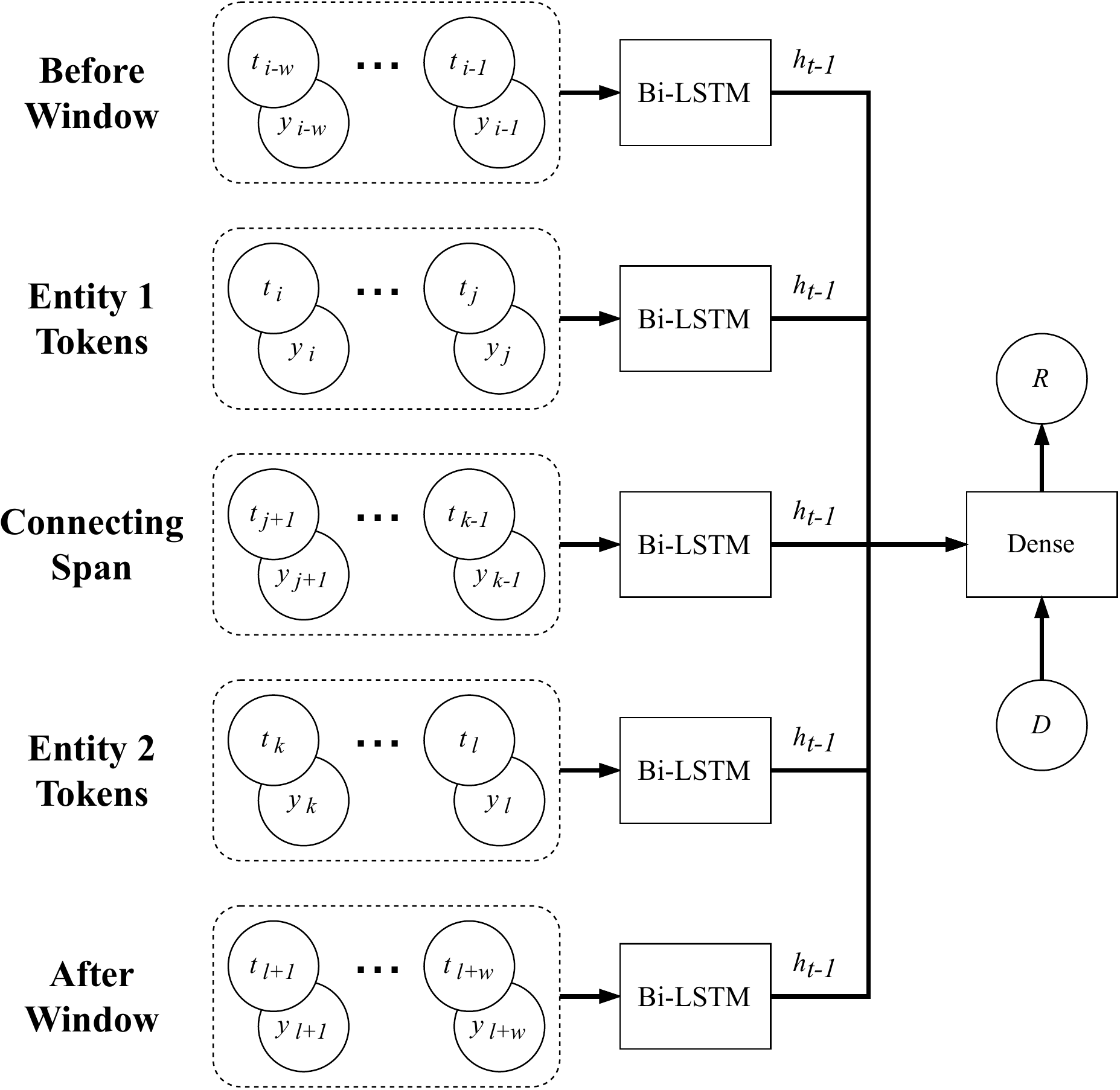}
	\caption{A schematic diagram of the neural Relation model. The Bi-LSTM nodes shown here refer to the same LSTM network, which is used for each of the spans. Here $t_n$ and $y_in$ indicate the token embedding and Entity label prediction for the $n^{th}$ token, respectively, $D$ is the direction bit indicating the direction of the Relation in the text, $R$ is the Relation prediction for this Entity pair and direction, and $w$ is the window width. The two Entities in question run from tokens $i$ to $j$, and tokens $k$ to $l$. The $h_{t-1}$ notations indicate that it is the hidden state from the final time-step which is used as the output from the LSTM nodes.}\label{fig:lstmrelmodel}
\end{figure}

A schematic diagram of this model is shown in Fig.~\ref{fig:lstmrelmodel}. The token embeddings in each of the five spans are concatenated with their BIO token label, and each span is passed through the same bidirectional LSTM network. The hidden state of the LSTM at the final timestep is used as a fixed-length representation of the span, and these five vectors are concatenated, along with the direction bit and the Entity labels for the start and end points of the proposed Relation (as one-hot encodings), and passed through a final dense output layer.

As with our Entity models, we use a trainable projection matrix to increase the model's capacity, and zero-pad the document to account for the windowed area.

\subsubsection{Rules-Based Model}

It is also useful to produce a rule-based model as a baseline for this Relation Classification task, in order to determine if a more complex trained model is justified -- as certain tasks are sufficiently tractable to be solved by much simpler models.
Hence this model is hand-crafted from observations of the available data to produce a robust set of rules which can predict Relations between labelled Entities in a document (as opposed to the trained statistical models previously discussed). By creating a heuristic model such as this, we allow ourselves to determine a baseline for model performance based on human intuition and knowledge of the domain. Without such a baseline, we have no way of contextualising model performance against a more easily interpretable algorithm.

This model uses two primary approaches: searching for patterns in the text between the two Entities (only practical for very short distance Relations), and using the patterns of Entities within sentences (ignoring individual tokens) to propose Relations which may exist between them.

For example, for examining the text between Entities, if we have a \ann{ParameterSymbol} annotation which is followed by a \ann{MeasuredValue} annotation, and the span of text between these two Entities is ``='' (ignoring any whitespace which may exist between them), then we can safely assume that the measurement is related to the symbol by a \ann{Measurement} Relation. There are other obvious connecting strings, such as ``\textbackslash{}sim'' or ``\textbackslash{}approx'', and similar strings for other Entity type pairings (e.g. ``<'' and ``>'' for \ann{ParameterSymbol} and \ann{Constraint} Entities).

However, this is insufficient for more complex sentences. For example, if an author is reporting multiple possible values for a physical parameter (e.g. dependant on different physical assumptions), then they may write a sentence of the form: ``If we make assumption X, we find a value for $A$ of 1.5, yet including assumption Y we find a value of 2.0.'' We observe that this pattern of \ann{ParameterSymbol} followed by multiple \ann{MeasuredValue} Entities is quite common, and so we can search for sentences which contain this pattern of Entity annotations, without needing to consider the constituent tokens (i.e. ignoring the textual content, and using only the order of Entities in the sentence). Similarly, a sentence which contains multiple measurements will often have a single \ann{ConfidenceLimit} Entity after all the values have been stated. Hence, we assume that any sequence of \ann{MeasuredValue} Entities followed by a single \ann{ConfidenceLimit} Entity can be linked such that each \ann{MeasuredValue} is connected by a \ann{Confidence} Relation to that \ann{ConfidenceLimit}.

A full list of the rules and patterns used for this model may be found in Appendix~\ref{app:rulesreldetails}.

\subsubsection{Relation Model Results}\label{sec:relationModelResults}

Table~\ref{tab:relclassresults} show the results of the top-performing model from our model search (performed as a grid search over model hyperparameters), along with the corresponding performance from our rule-based model. Model performance was again judged using the global F1 score calculated on the holdout test data.

The best performing neural model had an F1 score of 0.976, compared with 0.977 for the rule-based model. However, the similarity of these results is misleading, due to the heavy class imbalance in favour of the ``none`` label (due to the large number of possible Entity pairings). If we examine the per-class performance metrics for the models, we can see that the neural model suffers significantly in comparison to the rule-based approach, only achieving superior performance for the \ann{Measurement} Relation. From observation we find that the neural model struggles significantly with anything but the shortest Relations, where the Entities are very close to one another in the text, separated by only a few tokens.
However, the rule-based model shows good performance across the desired Relation labels, and so we shall be utilising this model for our final processing.

For the rule-based approach, the biggest issue remains the \ann{Property} Relation. This Relation is by far the most long-distance, often covering nearly the entire span of the text. As we are dealing with article abstracts here, it is common to have an object referenced at the beginning of the text, often in the first sentence (e.g. ``We examine the supernova SN 1998bu...''), followed by a description of the experimental approach, and then finally a concluding sentence stating the final result (``We find a peak luminosity of...''). This long-distance nature negates much of the sentence-level pattern matching we have leveraged for the rule-based approach. Additionally, if multiple celestial objects are mentioned in this way, or with some other oblique reference later in the text, it can be hard to distinguish which measurement belongs to which object using simple patterns. As such, the required simplifying assumptions produce a very low quality of predictions for the \ann{Property} Relation.

\begin{table}
\caption{Per-label performance metrics for the neural and rule-based Relation models. The values for the neural model are taken from the top-performing model from the model search.}\label{tab:relclassresults}
\centering
\begin{tabular}{cccccr}
\hline
\textbf{Model} & \textbf{Relation Type} & \textbf{Precision} & \textbf{Recall} & \textbf{F1} & \textbf{Support} \\
\hline
\multirow{5}{*}{Neural} & Confidence & .00 & .00 & .00 & 75\\
 & Measurement & .92 & .80 & \textbf{.86} & 655 \\
 & Name & .89 & .42 & .57 & 225 \\
 & Property & .00 & .00 & .00 & 159 \\
 & None & .97 & 1.00 & .98 & 21,807 \\
 \hline
\multirow{5}{*}{Rules} & Confidence & .86 & .80 & \textbf{.83} & 75 \\
 & Measurement & .93 & .75 & .83 & 655 \\
 & Name & .88 & .81 & \textbf{.84} & 225 \\
 & Property & .23 & .21 & \textbf{.22} & 159 \\
 & None & .98 & 1.00 & \textbf{.99} & 21,807 \\
 \hline
\end{tabular}
\end{table}

\subsection{Attribute Models}

For predicting Attributes we are considering only one model architecture, due to the relative simplicity of the problem. For this model we are only predicting Attributes relating to \ann{Constraint} values (\ann{LowerBound} and \ann{UpperBound}), due to the relative sparsity of the other Attribute labels in the training set, and so we only consider \ann{MeasuredValue} Entities when making predictions (as \ann{Constraint} values are not directly predicted, but inferred from the presence of Attributes). For example, in ``...finding $x \leq 0.5$ for...'', we would assign a \ann{UpperBound} Attribute label to ``0.5''. Note that here each \ann{MeasuredValue} Entity is considered as an individual classification task.

\begin{figure}
    \centering
	\includegraphics[width=\columnwidth]{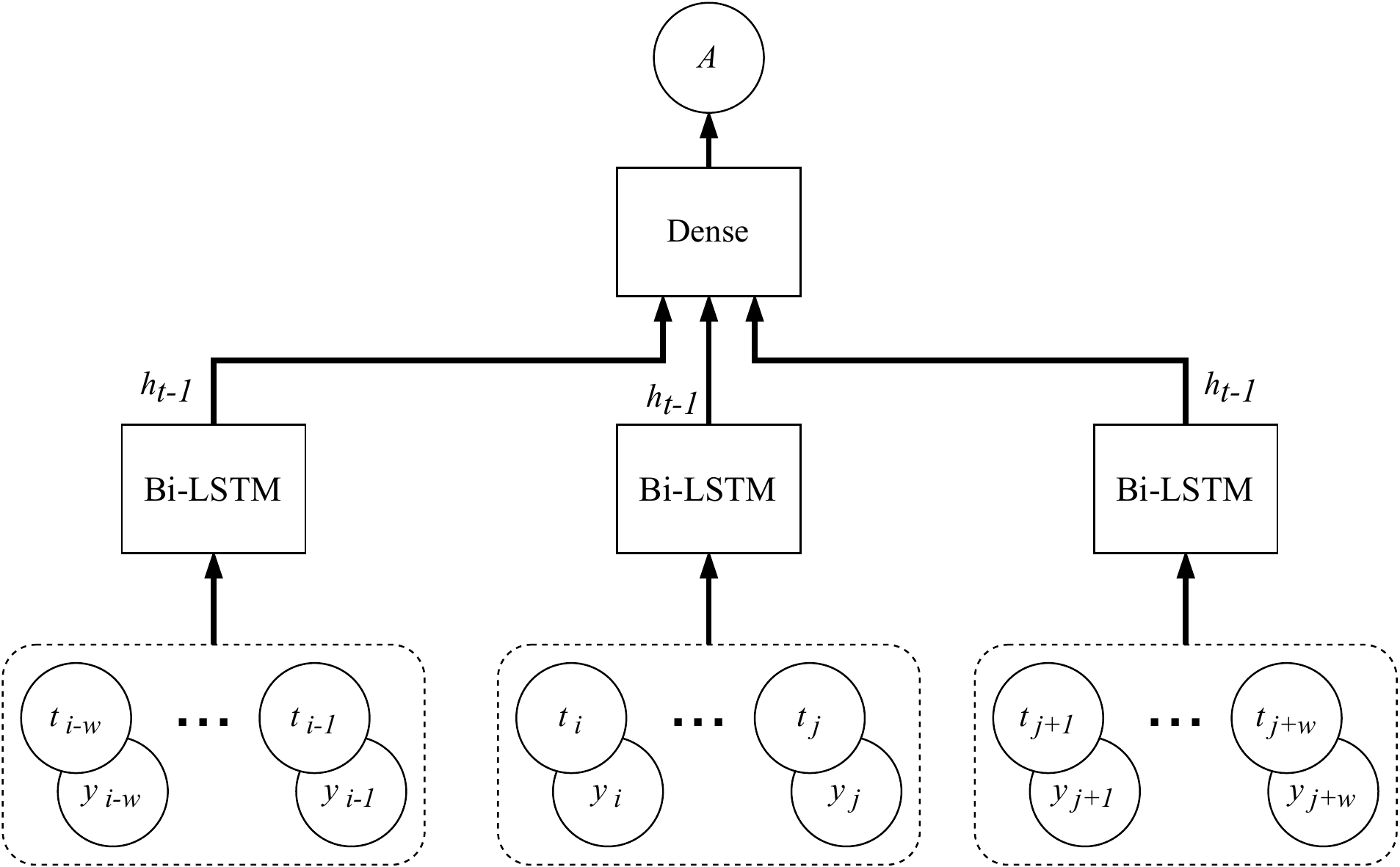}
	\caption{A schematic diagram of the Attribute model. The Bi-LSTM nodes here refer to the same LSTM network, passed over each span of tokens individually. Here $t_n$ and $y_n$ indicate the token embedding and model output for the $n^{th}$ token, respectively, $i$ and $j$ refer to index the start and end tokens for the Entity in question, and $A$ is the Attribute label prediction for that Entity. The $h_{t-1}$ notations indicate that it is the hidden state from the final time-step which is used as the output from the LSTM nodes.}\label{fig:attributemodel}
\end{figure}

A schematic diagram of this model is shown in Fig.~\ref{fig:attributemodel}. For this model we examine the tokens of the Entity itself, along with a fixed-width window around the Entity in question in both directions, and utilise a bidirectional LSTM layer to process these sequences of tokens. As for our Relation model, we use an LSTM to account for the variable length sequences we will encounter. The LSTM is used despite the fact that only the Entity token sequence is variable length (both window sequences are fixed length), as training on all sequences increases the training signal through the LSTM cells. As before, the Word2Vec embeddings are projected using a trainable projection matrix, and the predicted Entity label for each token is concatenated onto this projected embedding. The concatenated LSTM outputs (hidden state at final timestep) are then passed through a densely connected layer, producing the final output.

Using this model, we achieve the results shown in Table~\ref{tab:attclassresults}, using a grid-search over model hyperparameters. These correspond to an overall F1 score of 0.98. These results are considered to be of a reasonable quality to be used in our final pipeline.

\begin{table}
\caption{Per-label performance metrics for the top-performing Attribute model from the model search.}\label{tab:attclassresults}
\centering
\begin{tabular}{ccccr}
\hline
\textbf{Attribute Type} & \textbf{Precision} & \textbf{Recall} & \textbf{F1} & \textbf{Support} \\
\hline
LowerBound & .75 & .78 & .76 & 27 \\
UpperBound & .76 & .86 & .81 & 37 \\
None & 1.00 & .99 & 1.00 & 1939 \\
\hline
\end{tabular}
\end{table}

\subsection{Post-Processing}\label{sec:postprocessing}

With our models trained we combine their outputs and utilise them for prediction. For a given abstract, we first predict the presence of Entities in the text, by converting the token-level BIO Entity predictions into full Entity spans. This is done by simply identifying contiguous spans of tokens with the same predicted class, using ``begin'' tokens to identify the start of such sequences in cases where there are no separating \ann{Outside} tokens. If no ``begin'' token is present, the first ``inside'' token is assumed to be the beginning of the Entity. Next, any \ann{MeasuredValue} Entities are evaluated using the Attribute model to determine if they should be annotated with \ann{UpperBound} or \ann{LowerBound} Attributes, or simply left as \ann{MeasuredValue} annotations. If the Attribute model returns an appropriate prediction, the \ann{MeasuredValue} label is changed to a \ann{Constraint} label, with the appropriate bound Attribute. Finally, the Relation model is used to predict the presence of any Relations between the predicted Entities.

However, as is generally the case when dealing with natural language, the prediction outputs are not always as clean as we would desire -- especially in this context, where the textual entities we are searching for may be highly structured and brittle against minor errors (missing braces, for example). As such, post-processing steps are applied to the predictions before they are stored in a database, to remove obvious noise and false positives. Here we are dealing only with simple and glaring errors, rather than attempting to solve more subtle issues.

Full details of the post-processing steps applied to Entity and Relation annotations may be found in Appendix~\ref{app:postprocessing}. Note that no post-processing steps are applied to Attribute annotations (other than the Entity label replacement discussed previously).


\section{Results}\label{sec:results}

In this section we demonstrate a series of search queries on the model predictions for a variety of cosmological parameters. These will serve as examples of the kind of datasets which may be produced from these outputs.

\subsection{Hubble Constant}\label{sec:hubbleresultscomparison}

\subsubsection{Comparison with Rules-Based Model}

To begin our analysis of the processed neural model predictions, we compare the results
to that of our previous work in \citet{crossland1}, which utilised a rule-based approach
for identifying measurements based on a list of query strings.
This previous work focused on extracting measurements of the Hubble constant, $H_0$ -- chosen for this parameter's well-defined name and symbol, and the use of a commonly accepted standard unit for the quantity (km s$^{-1}$ Mpc$^{-1}$). The simplicity of the parameter identifiers was, essentially, a requirement of the approach, given that exact string matching was used in the algorithm. Our new approach should be capable of distinguishing all of the measurement patterns already identified in the rule-based approach, whilst also extending beyond these rigid (and hand-coded) patterns to encompass a more diverse range of writing styles.

For the rule-based model, we use the data from Fig.~4 in \citet{crossland1}, which used the following keyword strings for the search:
\begin{itemize}
\item Hubble constant
\item Hubble parameter
\item $H_0$: written `H\_0', `H\_\{0\}', `H\_o', `H\_\{o\}', `H\_\textbackslash circ', or `H\_\{\textbackslash circ\}'
\end{itemize}
For the neural model, we use a database of measurements created from the outputs of the final trained models from Section~\ref{sec:models}, and use the same keyword strings to extract measurement instances (note that the symbol normalisation discussed in Section~\ref{sec:postprocessing} will make some of the above symbol strings degenerate).

This produces datasets as follows: 2228 data-points for the rule-based model, and 872 data-points for the neural model.

After this initial search, both datasets have some additional constraints placed on them: 
\begin{enumerate}
    \item \label{itm:h0unit} We require that the measurement have units compatible with km s$^{-1}$ Mpc$^{-1}$. This leaves 584 and 578 data-points for the rule-based and neural models, respectively.
    \item \label{itm:h0unc} We require that the measurement have a stated uncertainty, or (in the case of the neural model) be a constraint value. This has the effect of reducing noise in the result set, and removing assumed or literature values, which are often reported without an accompanying uncertainty. 
\end{enumerate}

\begin{figure*}
    \subfloat[Rules-Based\label{fig:hubblekeyword}]{
        \includegraphics[width=0.45\textwidth]{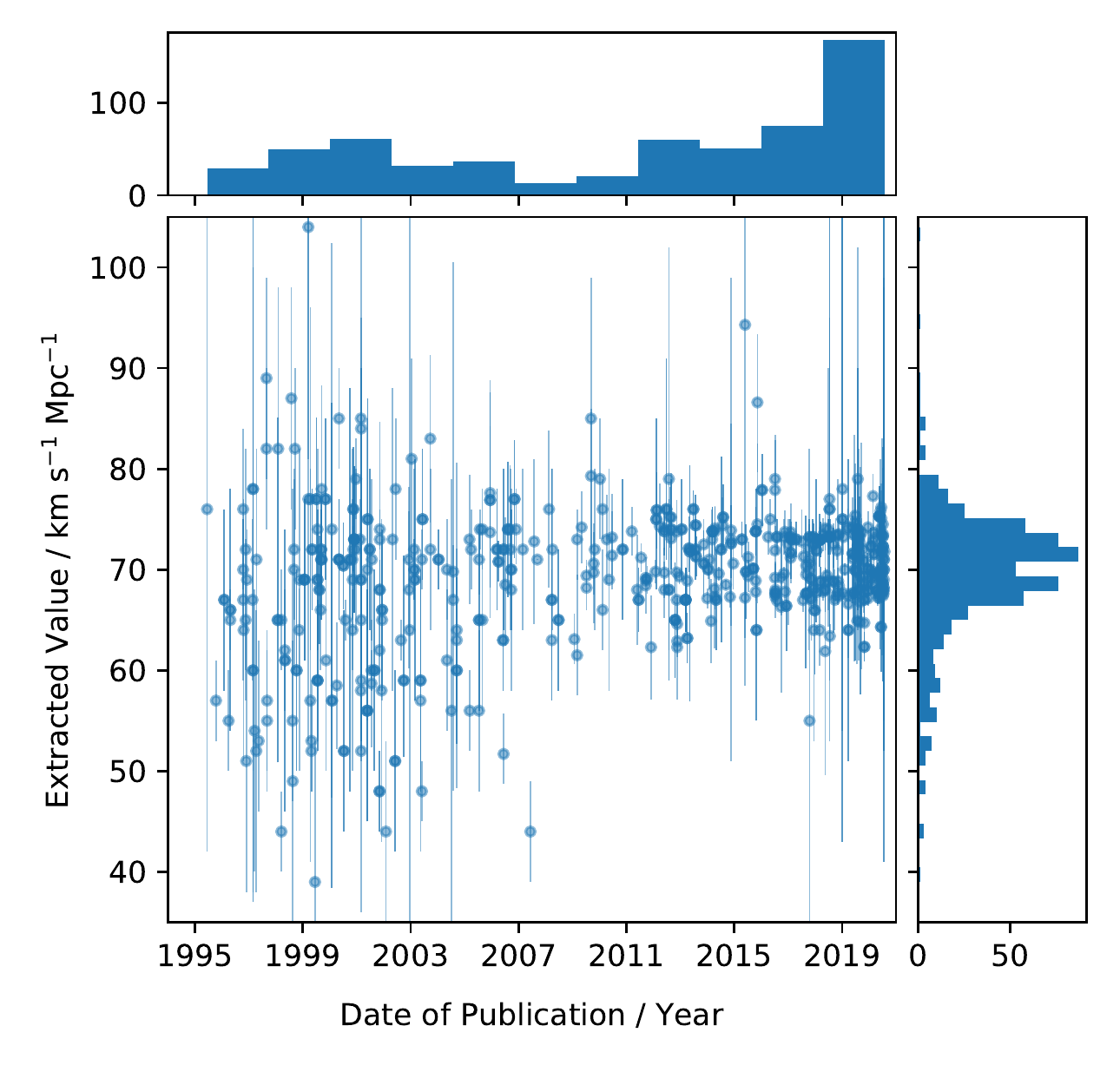}
    }
    \hfill
    \subfloat[Neural\label{fig:hubbleneural}]{
        \includegraphics[width=0.45\textwidth]{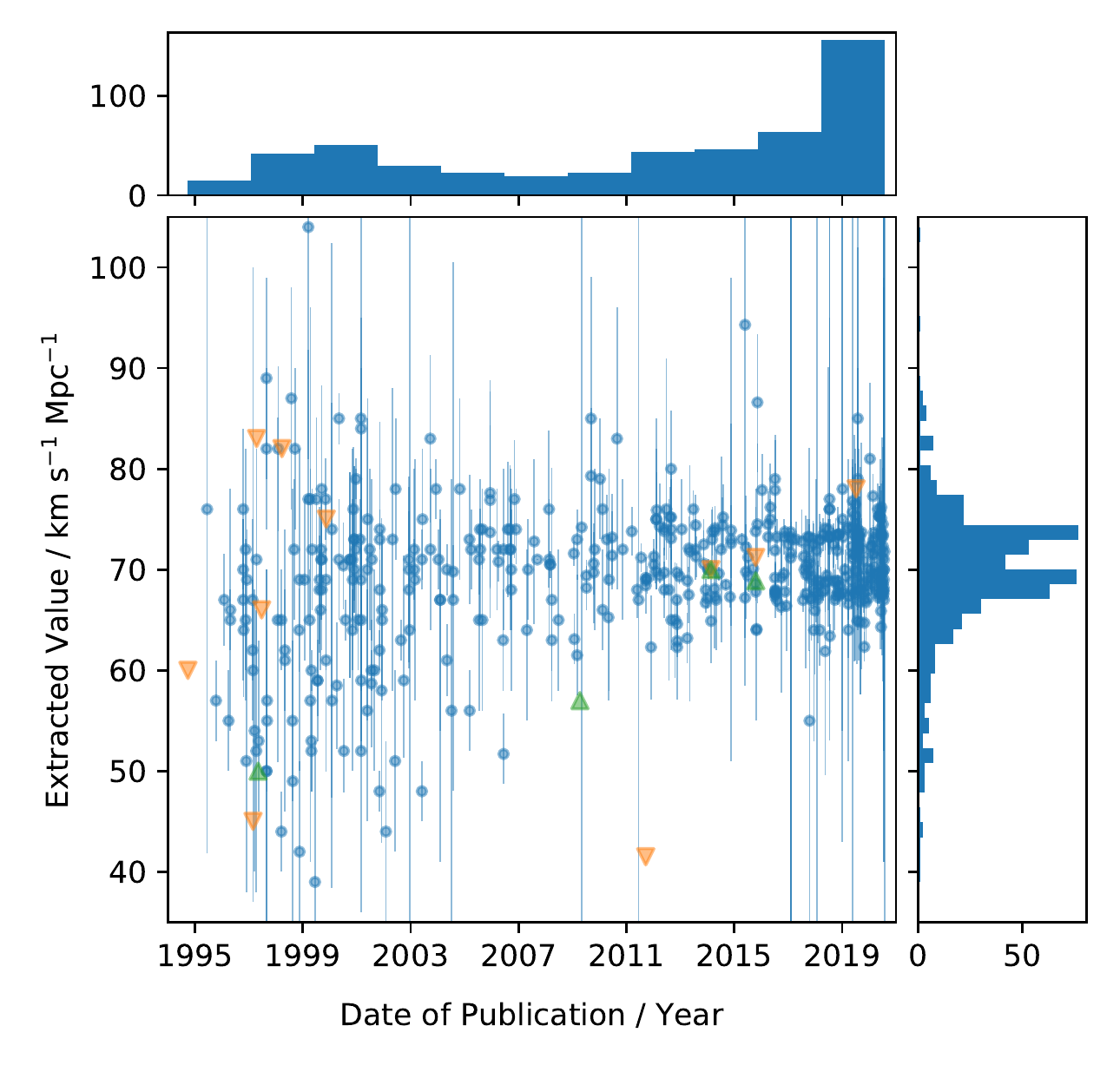}
    }
    \caption{Comparison of search results for the Hubble constant, $H_0$, from the rule-based (a) and neural models (b). In addition to the measurements provided as central values with stated uncertainties (i.e. ``$x \pm y$'', shown as blue circles with error bars), the neural model figure also shows values given in the source text as constraints (i.e. $H_0 < x$, or similar, shown as green arrows for lower bounds and orange arrows for upper bounds.)}\label{fig:hubbleoutputs}
\end{figure*}

This leaves us with the following datasets: 299 samples from the rule-based model, and 314 samples from the neural models, all with the correct units and a provided uncertainty or bound.
The outputs of the models are displayed as time-series (by publication date) in Fig.~\ref{fig:hubbleoutputs}.

From the effects of these cuts on the number of returned data-points we observe the following: The neural model is far more selective when identifying potential measurements in the text, finding far fewer potential spans initially. However, the identified spans are shown to be more grammatically relevant to the query phrases (``Hubble constant'', ``$H_0$'', etc.), given that a higher proportion survive our selection cuts using our existing knowledge of the Hubble constant (i.e. unit and required uncertainty): 13\% for the rule-based model versus 36\% for the neural model.

With these data collected and cross-referenced, we find an overlap of 261 samples, with 39 samples identified by the rule-based model that the neural model did not recover, and likewise 53 samples that only the neural model found. Most interesting out of these samples are the instances where only one model identifies a measurement, as they highlight gaps in the models' comprehension. To investigate this further the textual spans for both datasets were manually examined, and the following recurring failure states are noted (the examples reference those found in Table~\ref{tab:examples}):
\begin{enumerate}
    \item As seen in \citet{crossland1}, the rule-based model fails on a number of trivial cases, such as the presence of additional, unrelated numbers in the text, such as Example~\ExampleHzBreak{}, or more verbose language causing separation of keyword and measurement (as the model selects the closest measurement in the same sentence by character-distance). Many of these cases can be caught by the neural model -- however, long distance and multi-sentence Relations continue to pose a problem for both models.
    \item The rule-based model cannot distinguish standalone symbols from symbols as part of a larger span (indeed, no distinction is made in the keyword list between names and symbols at all). As such, it may misidentify instances of symbol search strings inside compound symbols, such as in Example~\ExampleHubbleAge{}. The neural model, however, looks at all tokens in context, and is not limited to a fixed set of symbols, and so will (ideally) identify the whole symbol span, as in Example~\ExampleHubbleAge{} where it correctly identifies the full symbol (therefore not returning the \ann{MeasuredValue} for our Hubble constant search, which specifies ``H \_ \{ 0 \}'' rather than ``H \_ \{ 0 \} \textasciicircum{} \{ -1 \}''). 
    \item Stray \LaTeX{} macros or other typographical anomalies can cause the regular expression patterns used by the rule-based model to miss potential measurements in the text, as for the failure in Example~\ExampleMathrmBreak{} for the rule-based model, where the unit string has been missed (and so the measurement is incomplete). The neural model, however, is more robust to these \LaTeX{} irregularities, and successfully annotated Example~\ExampleMathrmBreak{}.
    \item However, the neural model does stumble on certain styles of measurement reporting, most notably on brackets (``( )'') present in the middle of both measurements and symbols, as in Example~\ExampleOrTextBreak{}. This confusion is understandable, given that brackets often denote the beginning or end of an Entity annotation, and hence we can expect the model to be biased toward classifying bracket tokens as \ann{None} tokens (i.e. not belonging to any class), or transition from a run of tokens of one Entity type to another. This can either cause an Entity annotation to be incomplete, missing important tokens at the beginning and/or end, or split into multiple such incomplete annotations. For instance, in Examples~\ExampleOrTextBreak{} (Neural Model) \& \ExampleRandomBreak{} the \ann{MeasuredValue} Entity spans should be single \ann{MeasuredValue} annotations, but have been incorrectly identified as two separate spans due to the ``( or'' and ``( random'' tokens being labelled \ann{None} by the model. This means that, whilst having the correct token labels, the two Entity spans cannot represent the actual value of the measurement.
    \item \label{itm:neuralexactmatch} A notable point of failure for the neural results is the manner in which symbols are currently matched in the database: namely by using an exact match against the normalised symbol string (see Section~\ref{sec:postprocessing}). This leads to accurate annotations being ignored in our query in cases where a slight variation on the standard symbol has been used. An example of this can be seen in Example~\ExampleHoEPM{}, where the symbol ``H \_ \{ 0 \} \{ ( EPM ) \}'' has been correctly classified (as the bracketed portion was, presumably, intended as part of the symbol by the author -- here describing a methodology for the measurement), but does not exactly match the query string ``H \_ \{ 0 \}``. 
    \item Finally, the neural model suffers more broadly from uncertain classification of tokens at the beginning and end of Entities, commonly resulting in one or two missing or added tokens. Especially in the case of braces, where incomplete braces present a non-trivial post-processing issue, this can have a serious impact on parsing of symbols and measurements. This is especially true for symbols, where braces can imply sophisticated mathematical relations in composed symbols. This can be seen in Example~\ExampleHzBreak{}, where the \ann{ParameterSymbol} text contains unbalanced braces (as the numerical value has been incorrectly labelled as a \ann{MeasuredValue}).
\end{enumerate}

\begin{table*}
\centering
\caption{Example annotations from the rule-based model (``keyword'') and neural model. The rule-based model does not use an annotation schema, and so the identified spans have been simply labelled ``Keyword'' or ``Measurement'', whereas the examples from the neural model use the annotation labels from Section~\ref{sec:annotationproject}.}
\label{tab:examples}
\begin{tabular}{ccl}
\hline
Number & arXiv Identifier & Tokenized and Annotated \TeX{} Source \\
\hline
\ExampleHzBreak & 1311.1767 &
\begin{tabular}{cc}
    Keyword Model: & \includegraphics[scale=0.45]{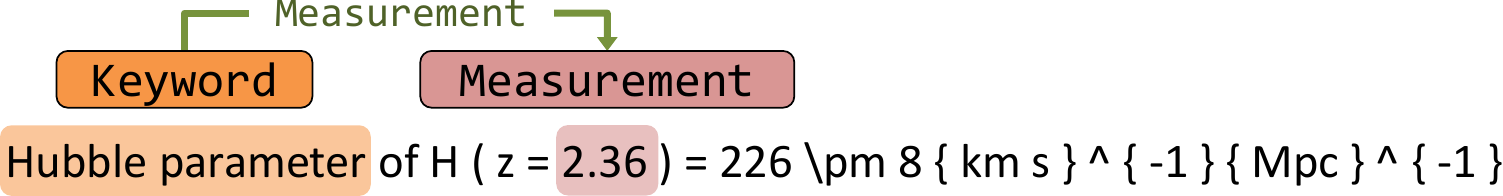} \\[10pt]
    Neural Model: & \includegraphics[scale=0.45]{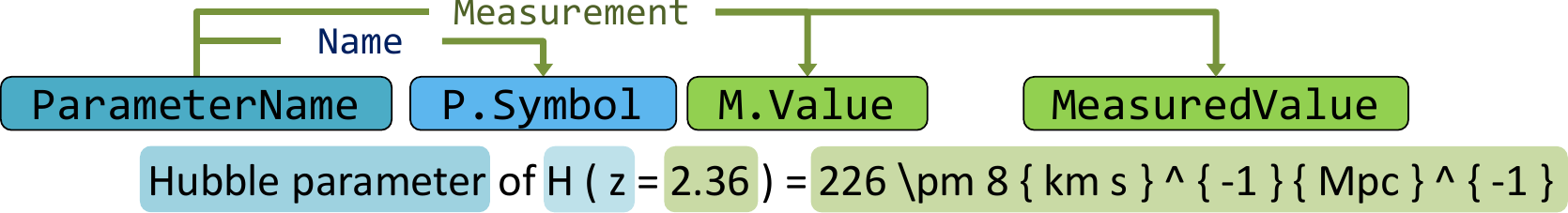}
\end{tabular}  \\ \hline
\ExampleHubbleAge  & 0704.3267 &
\begin{tabular}{cc}
    Keyword Model: & \includegraphics[scale=0.45]{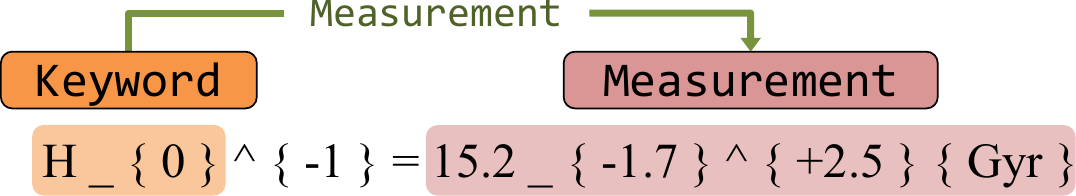} \\[10pt]
    Neural Model: & \includegraphics[scale=0.45]{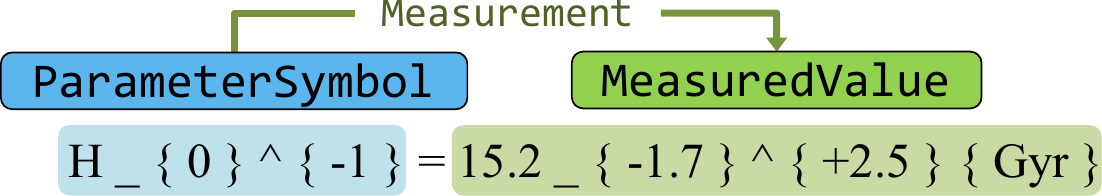}
\end{tabular}  \\ \hline
\ExampleMathrmBreak & 1005.0263 &
\begin{tabular}{cc}
    Keyword Model: & \includegraphics[scale=0.45]{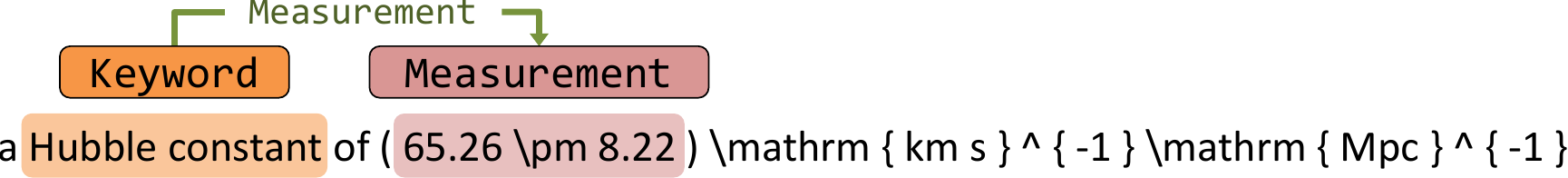} \\[10pt]
    Neural Model: & \includegraphics[scale=0.45]{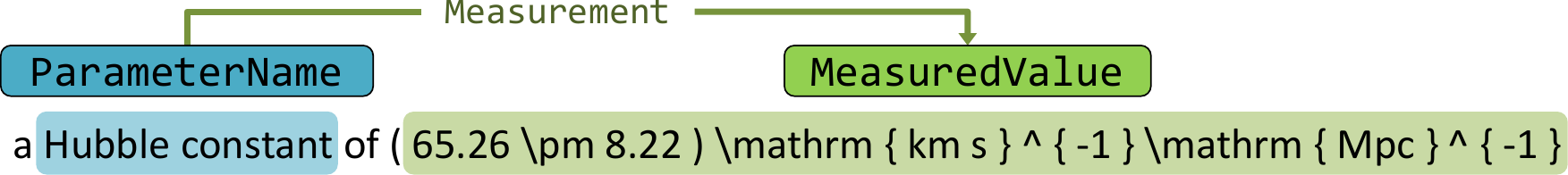}
\end{tabular}  \\ \hline
\ExampleOrTextBreak & 1105.5206 &
\begin{tabular}{cc}
    Keyword Model: & \includegraphics[scale=0.45]{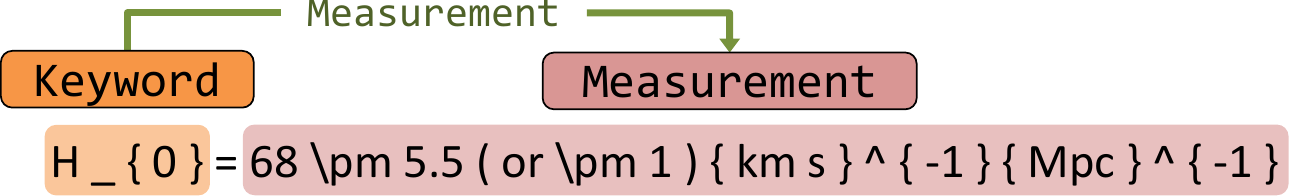} \\[10pt]
    Neural Model: & \includegraphics[scale=0.45]{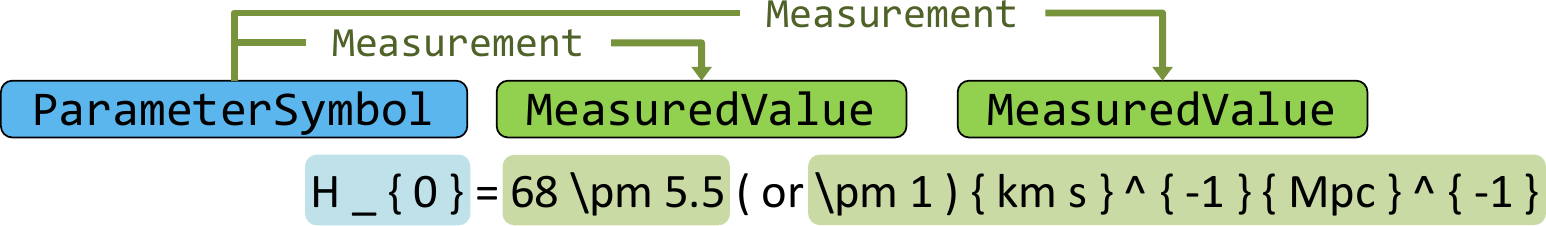}
\end{tabular}  \\ \hline
\ExampleRandomBreak & 1403.1693 &
\begin{tabular}{cc}
    Neural Model: & \includegraphics[scale=0.45]{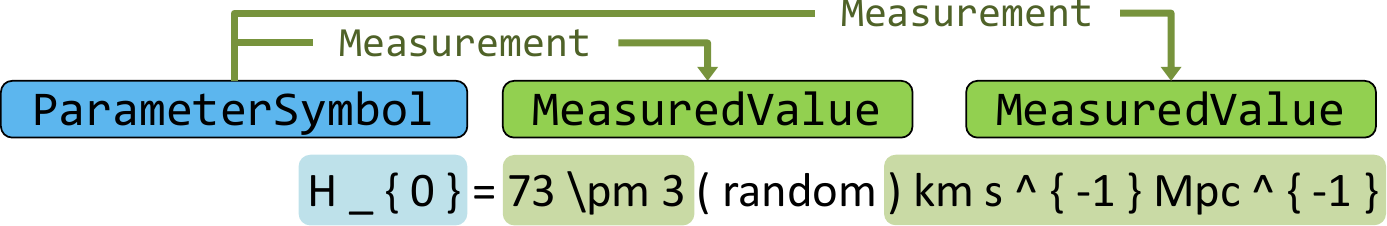}
\end{tabular}  \\ \hline
\ExampleHoEPM & astro-ph/0305259 &
\begin{tabular}{cc}
    Keyword Model: & \includegraphics[scale=0.45]{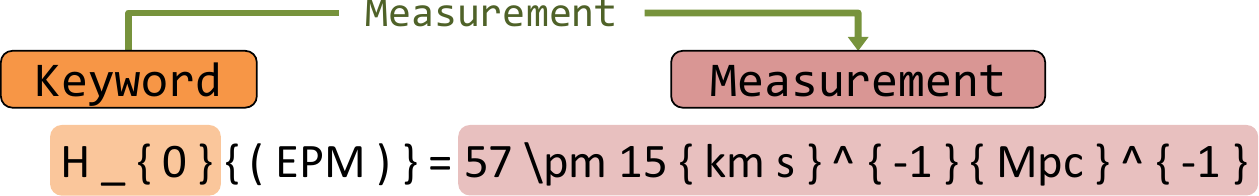} \\[10pt]
    Neural Model: & \includegraphics[scale=0.45]{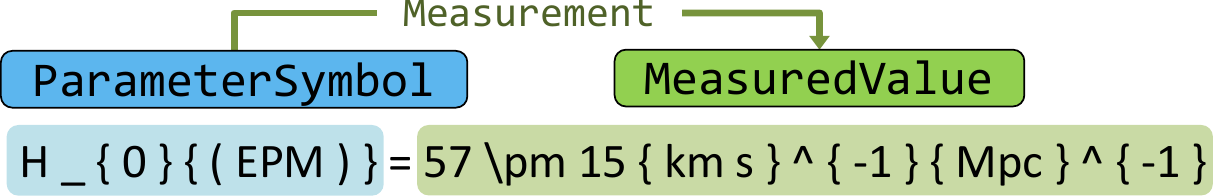}
\end{tabular}  \\ \hline
\end{tabular}
\end{table*}

From these observations, and the results of our comparison of the model outputs, we conclude that the neural model is capable of catching the large majority of cases covered by the rule-based model, and has the capacity to distinguish far more complex linguistic and typographical patterns than the rigid rule-based approach by considering token context.
However, manual examination of the model outputs shows that the neural model also suffers from incorrect classification of Entities, resulting in similar problems to those seen in \citet{crossland1}. As such, we have not yet moved beyond the requirement for some prior knowledge from the user to filter and refine the search results.

\subsubsection{Discussion}

From the collected data shown in Fig.~\ref{fig:hubbleoutputs}, we may also note the presence of certain trends in the measurement values of the Hubble constant. Of particular interest is the spike in reported measurements over the last few years, which could not be seen in the dataset used previously. This is thought to be due to the high profile tension which has arisen in recent years between the early and late universe determinations of the Hubble constant. Measurements based on the early universe, notably measurements from the CMB by the \textit{Planck Mission} \citep{Planck2018}, give a consistently lower value for $H_0$, approximately 67 km s$^{-1}$ Mpc$^{-1}$. Whereas late universe measurements, generally using standard candles such as Cepheids and Type Ia supernovae (and other, more novel objects, such as miras, masers, lensing objects), lead to values slightly above 70 km s$^{-1}$ Mpc$^{-1}$ - these measurements also having become more prevalent lately, with the release of data from projects such as the \textit{Gaia Mission} \citep{GaiaMission}.

Over the last decade the measurement uncertainties on values for the Hubble constant have been decreasing (as can be seen in Fig.~\ref{fig:hubbleoutputs}), and with the publication of the results from the \citet{Planck2018} the $>3\sigma$ tension between these two epochs has become the topic of much debate \citep{RiessHubbleTensionSummary}. In our results here we may see this narrative unfold, from the decreasing uncertainties through to the explosion in the number of reported measurements after 2018 (see the time-axis histograms in Fig.~\ref{fig:hubbleoutputs}). This tension may be clearly seen in our model outputs\footnote{The query to reproduce the data from Fig.~\ref{fig:hubbleoutputs} may be found at: \url{http://numericalatlas.cs.ucl.ac.uk/constant/hubbleconstant}} from the two distinct peaks in the distribution of $H_0$ values in Fig.~\ref{fig:hubbleoutputs} (see vertical axis histograms).

\begin{figure}
    \centering
	\includegraphics[width=0.9\columnwidth]{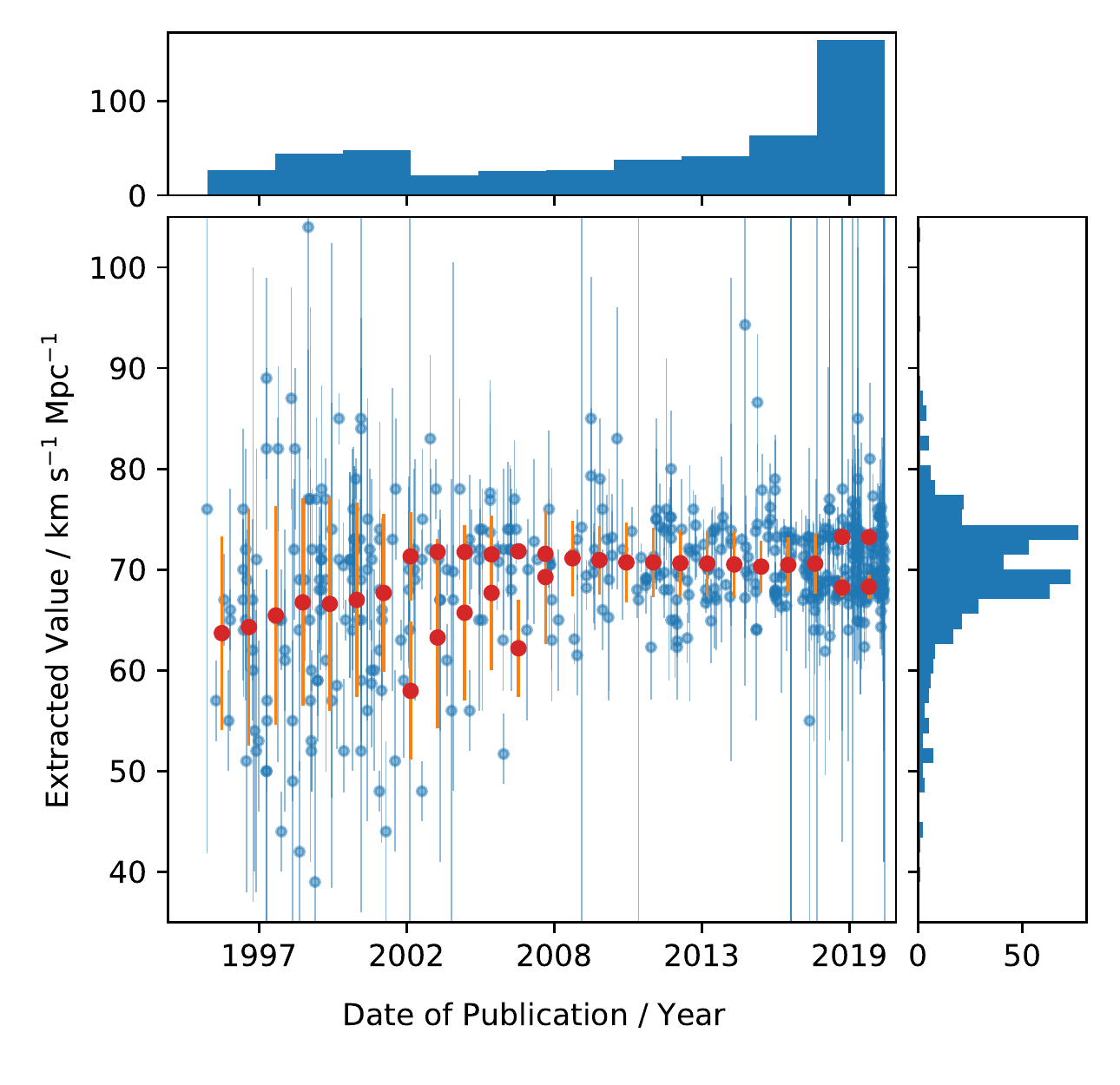}
	\caption{Time series of the search results for $H_0$, showing reported value against publication date (blue), along with the mean (red points) and dispersion (error bars) of the fitted Gaussian distributions for overlapping 5-year periods.}\label{fig:h0xdfit}
\end{figure}

In order to better visualise the changing understanding of $H_0$ we have used the Extreme-Deconvolution (XD) algorithm \citep{xdPaper} to fit Gaussian mixture models on overlapping 5-year bins of the search results, as shown in Fig.~\ref{fig:h0xdfit}. This algorithm uses the stated uncertainties of the measurements in the fit, giving a better representation of the consensus value in the considered period. The Akaike information criterion \citep{aicPaper} is used to determine the optimal number of components for the mixture models.
From these fits we clearly observe the decreasing measurement uncertainty in $H_0$ over time, followed by the bifurcation in the distributions after the Planck results.

We also see additional interesting features, such as the sudden increase in high-uncertainty values reported during this recent spike in popularity. Examination of these papers shows that this is due to various novel experimental techniques being explored in order to resolve the tension, such as: gamma-ray burst supernovae \citep{1805.06892}, AGNs \citep{1903.12308, 1906.08417}, Luminous Red Galaxies \citep{2005.13126}, and lensing objects \citep{2007.02941, 2007.14398}. There is also a notable number of uses of gravitational wave signals \citep{1806.10596, 1807.05667, 1901.01540, 1909.00587, 1909.09609, 2006.14961, 2007.11148} to determine values for the Hubble constant. We can see that, in addition to the raw numerical values returned by our search, there are rich possibilities with these data for analysis of uptake of ideas and techniques within the astrophysics community.

\subsection{Application to Other Cosmological Parameters}\label{sec:resultscosmoconstants}

Having shown that our new model can perform well compared to our baseline on a well-structured case, we move on to more challenging examples. We note from our examination of the Hubble constant that filtering our result set by a known unit is a very effective way of identifying incorrect samples (especially for the Hubble constant, with a rather specific common expression for its dimensionality -- as opposed to something more generic, e.g. K or kpc). However, there are many interesting quantities with more common units -- or, indeed, dimensionless quantities.

However, the dimensionality filtering for the Hubble constant had far less impact on the result set from the neural model, with a drop off in samples of only 33\% for this step, in comparison to 74\% for the rule-based model. This suggests then, as noted previously, that the neural model is already far more selective when identifying measurement spans in the text, and hence relies less on post-processing to identify candidate measurements.

Furthermore, the availability of both a common, well-defined name and symbol for the Hubble constant is a special case in the scientific literature, and we must extend beyond this if we hope to produce a useful tool for the community.

We shall now present some test cases which emphasise this more challenging regime. For this, we have chosen a set of the cosmological parameters, as they are quantities of interest in the scientific community with uncertain values and a relatively large catalogue of reported measurements, which exhibit the challenging features mentioned above. We use the following list of parameters as case studies:
\begin{enumerate}
    \item \label{itm:omegam} $\Omega_M$, the ratio of the present matter density to the critical density,
    \item \label{itm:omegal} $\Omega_\Lambda$, the cosmological constant as a fraction of the critical density,
    \item \label{itm:sigma8} $\sigma_8$, the amplitude of mass fluctuations,
    \item \label{itm:omegab} $\Omega_b h^{2}$, the baryon density parameter,
    \item \label{itm:primspecidx} $n_{\textrm{s}}$, the primordial spectral index,
    \item \label{itm:mnu} $\sum m_\nu$, the sum of neutrino masses,
    \item \label{itm:omegak} $\Omega_k$, the curvature,
    \item \label{itm:darkenergyw} $w_0$, the equation of state parameter for dark energy.
\end{enumerate}
Fiducial values for each of these parameters may be found in Table~\ref{tab:fiducialvalues}, which correspond to those reported by \citet{Planck2018}.

\begin{table}
\caption{Fiducial values of the cosmological constants taken from \citet{Planck2018} for comparison with model results.}
\label{tab:fiducialvalues}
\centering
\begin{tabular}{ccc}
\hline
Parameter & Value & $1\sigma$ Error Bar \\ \hline
$\Omega_M$ & 0.315 & 0.007 \\
$\Omega_\Lambda$ & 0.6889 & 0.0056 \\
$\sigma_8$ & 0.811 & 0.006 \\
$\Omega_b h^2$ & 0.02242 & 0.00014 \\
$n$ & 0.965 & 0.004 \\
$\sum m_\nu$ & < 0.12 eV & -- \\
$\Omega_k$ & 0.001 & 0.002 \\
$w_0$ & -1.03 & 0.03 \\
\hline
\end{tabular}
\end{table}

These parameters present a variety of interesting challenges to our models:
    Many of the parameters in question lack a well defined name -- which is not to say that they do not have established naming conventions, but that these conventions present greater challenges than a moniker such as ``the Hubble constant''.
    For example, the word ``curvature'' is relatively generic, in the context of astrophysics at large, but in the context of a cosmology paper could reasonably be used with very little other explanatory information to refer to $\Omega_k$.
    So, something more specific would be required. Yet the phrase ``spatial curvature of the universe'' has large potential for linguistic variability.
    This means we require our model to be able to identify grammatically significant sequences of tokens in the text, rather than simple name-phrases like ``Hubble constant''.
    
    Additionally, many of the symbols for these parameters are commonly found in compound expressions, which proved a major stumbling block for our initial keyword search. For example, we wish to be able to distinguish between the Hubble age expressed as ``$H_{0}^{-1}$'', and the Hubble constant expressed as ``$H_{0}$'', or between expressions such as ``$\Omega_M$'' and ``$\Omega_{m} h^{2}$''. For this, once again, we require not just to find the tokens of interest, but to take account of their context in the sentence.
    
    Finally, the majority of these parameters are dimensionless. This presented a major hurdle to our previous approaches for identifying measurements, as filtering candidate spans by stated units was an important step in reducing noise in the result set.

With these parameters we will show both the power of our model, but also the utility of our framework and how it may be used to intelligently search through the collected data to find sets of measurements relating to a certain physical quantity.

\subsubsection{Matter Density Parameter, $\Omega_M$}

\begin{figure*}
    \subfloat[Neural\label{fig:omegamneural}]{
        \includegraphics[width=0.45\textwidth]{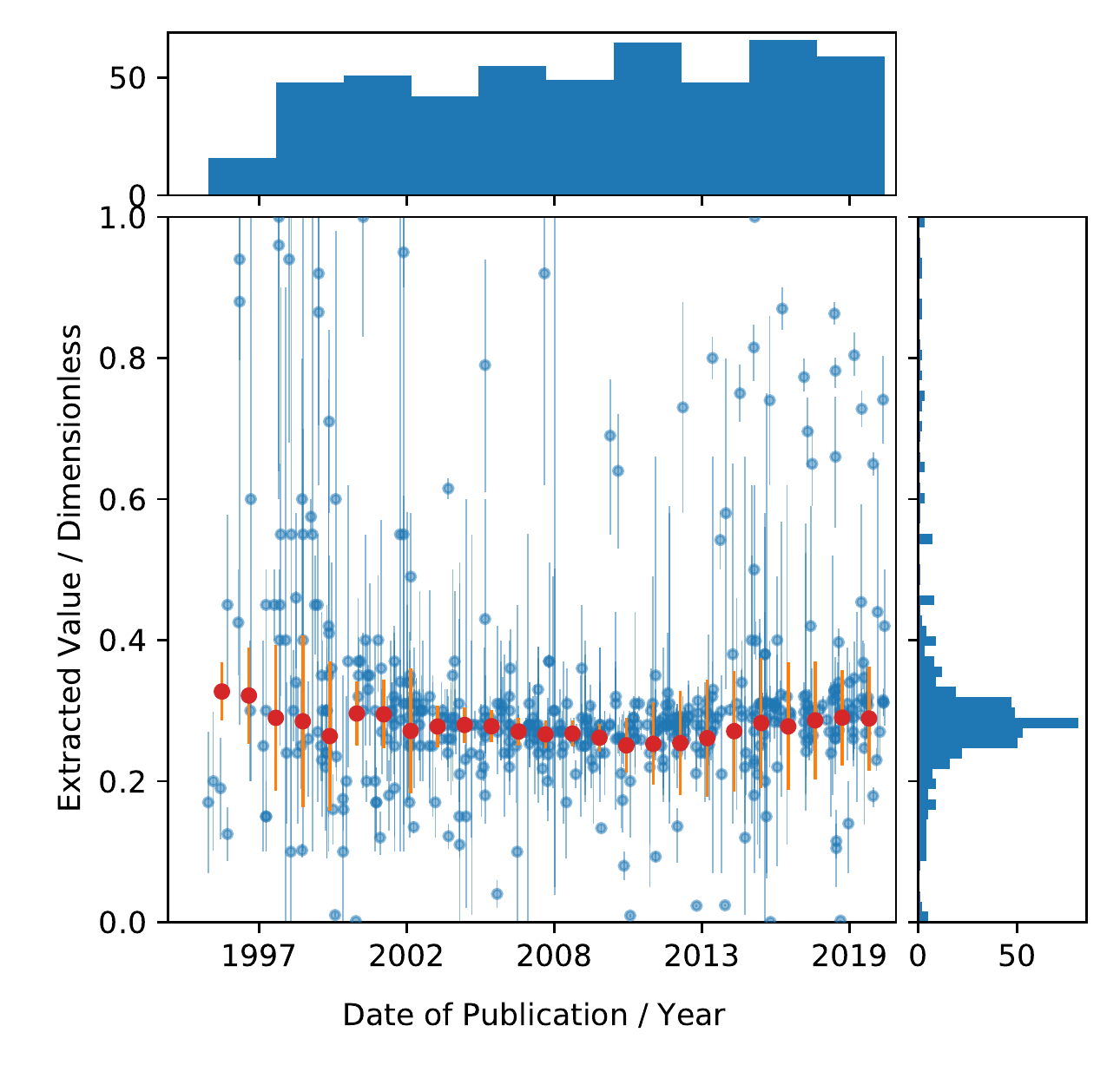}
    }
    \hfill
    \subfloat[Rules-Based\label{fig:omegamkeyword}]{
        \includegraphics[width=0.45\textwidth]{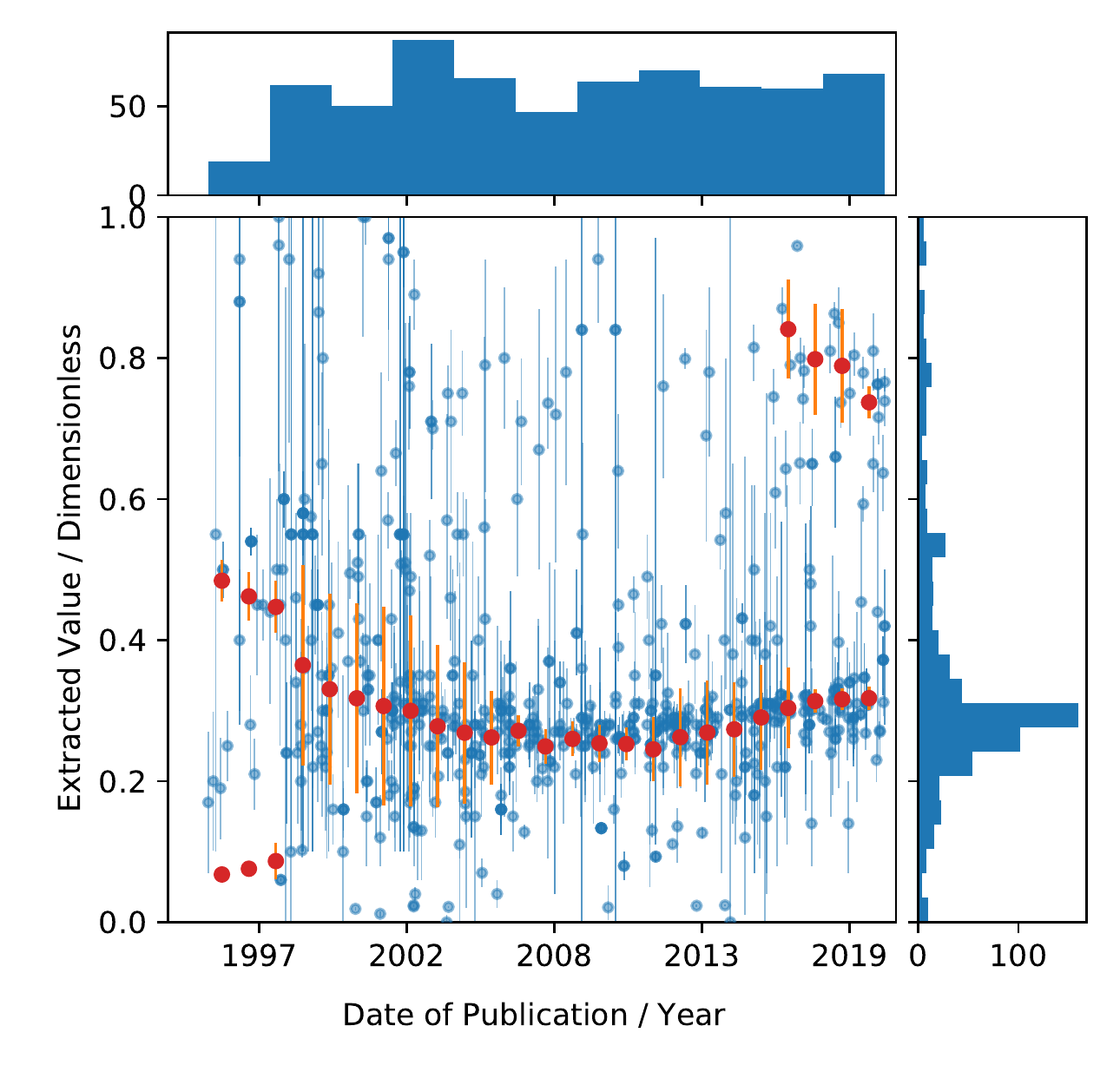}
    }
    \caption{Comparison of search results for the rule-based and neural models for the cosmological matter density, $\Omega_{M}$. Both plots show only measurements which report a central value and an uncertainty (the neural model also contains constraint measurements, but these have been omitted for clarity), shown in blue. The mean (red points) and dispersion (error bars) of the Gaussian mixture models fitted on overlapping 5-year bins are also shown.}\label{fig:omegam}
\end{figure*}

To begin, let us consider the matter density parameter, $\Omega_M$. Our search parameters are as follows\footnote{The query to reproduce the data in this plot may be found at: \url{http://numericalatlas.cs.ucl.ac.uk/constant/omegam}}:
\begin{itemize}
    \item Name: ``mass density'', ``matter density''
    \item Symbol: ``\textbackslash{}Omega \_ \{ M \}'', ``\textbackslash{}Omega \_ \{ m \}'', ``\textbackslash{}Omega \_ \{ 0 \}''
    \item Unit: Dimensionless
    \item Value range: $0 \leq x \leq 1$
\end{itemize}
From this query we find 1408 candidate measurements. Examination of the measurements and their associated names and symbols shows some false positives, for example ``baryonic mass density parameter'' and ``amplitude parameter of the matter density fluctuations'' being incorrectly identified using our inclusion-based string matching (for \ann{ParameterNames}). However, the large majority of cases display sensible name/symbol combinations. The mean value of parsed measurements is 0.385, with a median value of 0.3. If we now add the stipulation that measurements must provide an uncertainty to be included in the result set, we find 449 values with a mean value of 0.297 and a median of 0.28. A plot of these identified measurements (uncertainty required), by publication date, is shown in Fig.~\ref{fig:omegamneural}, along with Gaussian mixture models fitted using the XD algorithm (in the same manner as the $H_0$ plots). The figure shows a clear peak in the measurement distribution at a value of approximately 0.3, as expected from the known history of $\Omega_M$, and shows the varying trend in the community's measurements of the parameter over the last two decades. It should be noted that there is no distinction made in the search query or the plot between measurements which assume a spatially flat universe and those which do not.

For comparison we have also plotted the results of this same query using the rule-based model in Fig.~\ref{fig:omegamkeyword}. Whilst the same general trends are observed in both plots, there is a broader distribution of outliers visible in the rule-based results. This is clearly visible in the fitted distributions, which are much more confined for the neural results.
We also note that the neural model produces a smaller number of results overall (449 for the neural model versus 645 for the rule-based model), along with a mean value closer to the expected result (0.297 for the neural model versus 0.357 for the rule-based model). This further shows that the neural model has better intrinsic selectivity than the rule-based model, without the need for filtering based on dimensionality.

Fig.~\ref{fig:omegam} demonstrates the community's understanding of $\Omega_M$ over the last two decades. The most decisive event appears to be the WMAP results from the First \citep{WMAP2003} and Three-Year \citep{WMAP2006} data releases. The years following these landmark papers see a much more confined region for the proposed values of $\Omega_M$ than the preceeding years. This is especially true throughout the majority of 2004, where the publications present values with tighter constraints than in the surrounding years. Considering that these publications utilise different data sources and techniques -- including combinations of supernova and X-ray observations \citep{astro-ph/0403228, astro-ph/0404201}, large scale structure with supernovae data \citep{astro-ph/0405118}, Chandra observations of clusters \citep{astro-ph/0405340}, combining the integrated Sachs Wolfe effect and supernovae data \citep{astro-ph/0407022}, SDSS data \citep{astro-ph/0408003}
-- yet still find observations in such tight agreement, it is possible we are seeing a period of confirmation bias here. After the WMAP Three-Year data release however, we see a period of relatively stable values and constraints on the value of $\Omega_M$, which exhibits a slight trend towards increasing values over time. An exception to this is the 2014-16 period, where a number of observations with much larger uncertainties may be seen. The use of lensing data appears to be a contributing factor to these measurements \citep{1403.5278, 1407.3955, 1412.3683, 1512.04555} in addition to the innovative use of SDSS results, including the Alcock-Pacynski Test with Cosmic voids \citep{1602.06306}, and utilising HII regions as standard candles \citep{1608.02070}. Following this period, we once more see a return to a relatively stable understanding of the quantity, yet with more variation between reported measurement values (as shown by the fitted distributions), with a trend towards a slightly higher value over time -- following the trajectory from the ${\sim}0.281$ WMAP value \citep{WMAP2013} to the ${\sim}0.315$ value reported by \citet{Planck2018}.


Finally, it should be noted that the cluster of values at $0.7-0.8$ after 2010 are erroneous, and are almost all due to a misidentified \ann{ParameterSymbol} annotation involving the quantity $S_8 = \sigma_8 ( \Omega_M / 0.3 )^{0.5}$.

\subsubsection{Cosmological Constant Parameter, $\Omega_\Lambda$}
\begin{figure}
    \centering
	\includegraphics[width=0.9\columnwidth]{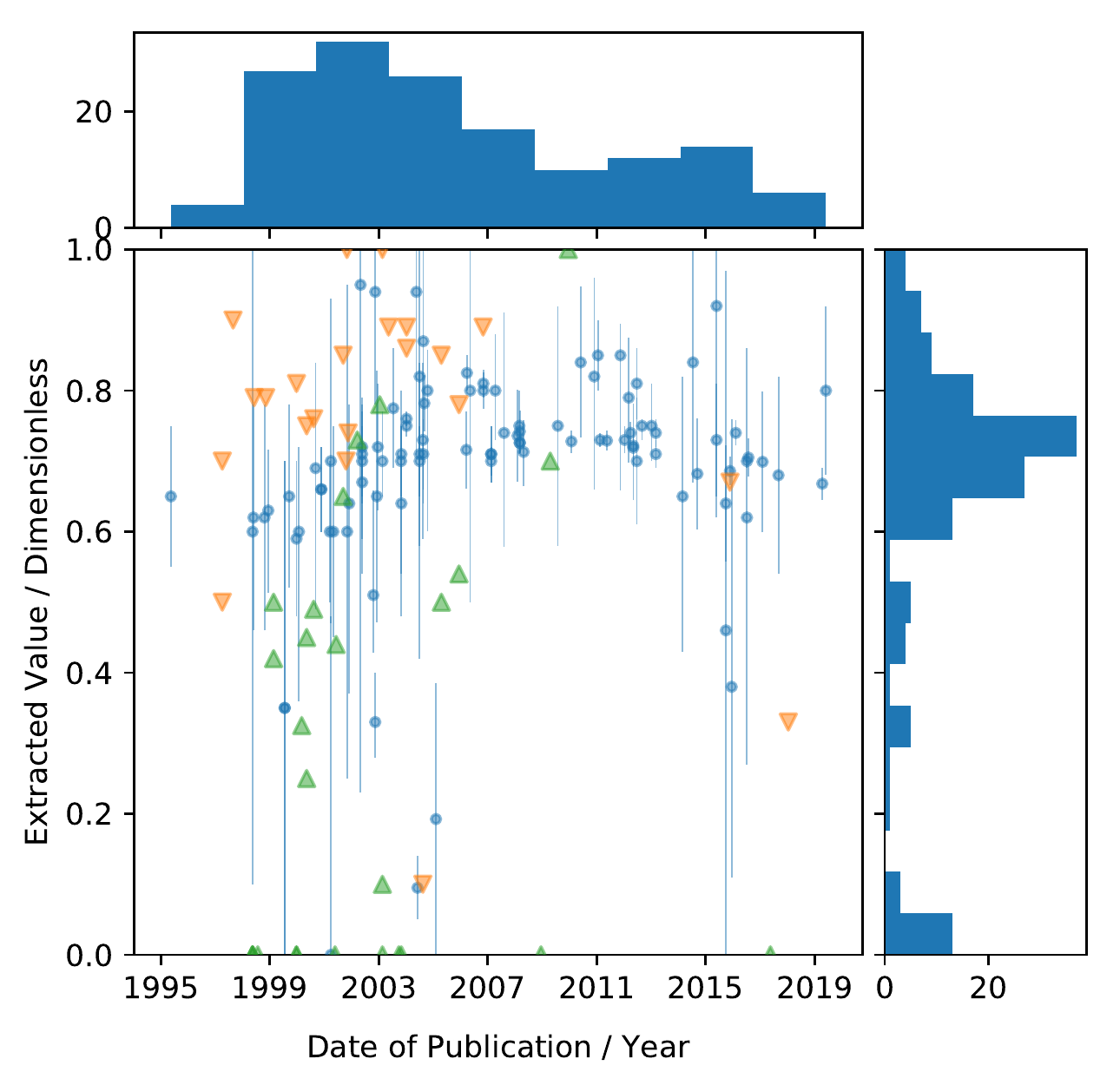}
	\caption{Time series of the search results for $\Omega_{\Lambda}$, showing reported value and publication date.}\label{fig:omegalambda}
\end{figure}

As a complement to our previous example, we examine the Cosmological Constant as fraction of critical density, $\Omega_\Lambda$. Here we use the following search parameters\footnote{The results of this query may be found at: \url{http://numericalatlas.cs.ucl.ac.uk/constant/omegalambda}}:
\begin{itemize}
    \item Symbol: ``\textbackslash{}Omega \_ \{ \textbackslash{}Lambda \}''
    \item Unit: Dimensionless
    \item Value range: $0 \leq x \leq 1$
\end{itemize}
We find 421 results, with a mean value of 0.592, and a median of 0.7.
Without requiring uncertainties, we find that more than half of the returned values are assumed values for the parameter (generally without provided uncertainties) clustered at the values 0.0 and 0.7.
The usage of these assumed values appears to drop off after 2004 for 0.0, and 2007 for 0.7. Requiring uncertainties, we find 88 values with a mean of 0.713 and a median of 0.712. A time-series plot of these measurements is shown in Fig.~\ref{fig:omegalambda} (again, no distinction is made in the search query between measurements reported assuming a spatially flat Universe and otherwise).

Here also we see trends in the community's understanding of this value: a particularly striking change is the drop off in values reported as upper or lower limits (i.e constraints) on $\Omega_\Lambda$ (e.g. ``$\Omega_\Lambda > 0.5$''), in favour of central values with uncertainties (e.g. ``$0.7 \pm 0.1$''), coinciding with the WMAP Three-Year Data Release \citep{WMAP2006}. It would appear that the influence of the WMAP data led to an acceptance of better constraints among the community, and hence a shift away from reporting $\Omega_\Lambda$ as a constraint. Additionally, we once again see an increase in measurement uncertainties during the 2014-16 period. The publications in question make use of galaxy cluster and quasar observations \citep{1402.6212, 1505.07118, 1512.04555, 1607.01790}, galaxy halo models \citep{1407.3811}, and gamma-ray bursts \citep{1509.08558}. Given the timing of these publications, it is quite possible that this additional debate around the value may be related to the release of the Planck 2015 results \citep{Planck2015} -- possibly both in preparation (or anticipation) as well as in response.

There is also an interesting value reported by \citet{astro-ph/9505066}, an early exploration of dark energy cosmology models using observational constraints. This publication appears to be several years ahead of the Nobel prize measurement of $\Omega_\Lambda$ \citep{1998Natur.391...51P, 1998ApJ...507...46S}, and has perhaps not received a proportional amount of attention.

\subsubsection{Amplitude of Mass Fluctuations, $\sigma_8$}
\begin{figure}
    \centering
	\includegraphics[width=0.9\columnwidth]{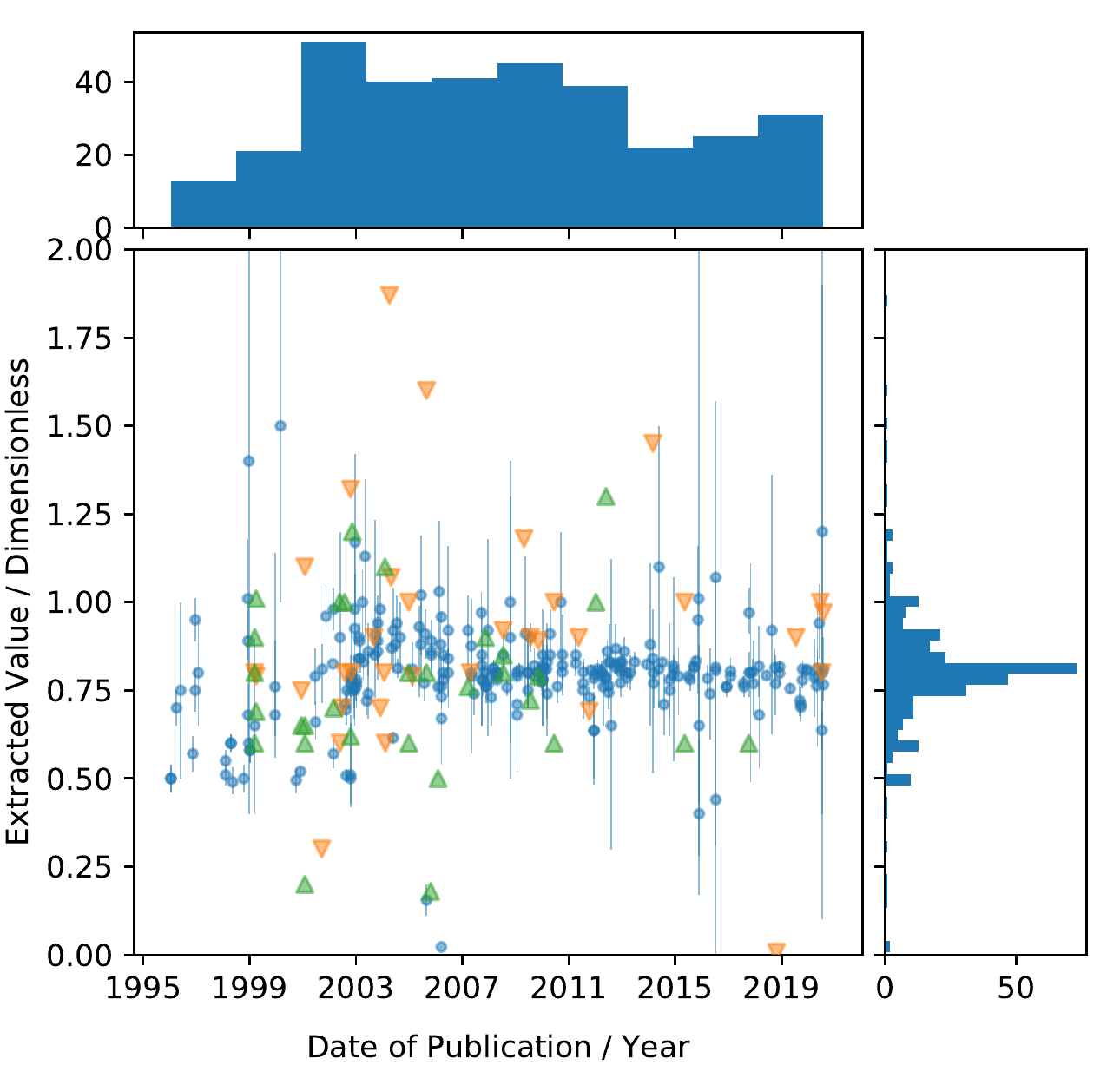}
	\caption{Time series of the search results for $\sigma_{8}$, showing reported value and publication date.}\label{fig:sigma8}
\end{figure}

Next we consider the amplitude of mass fluctuations, $\sigma_8$, with the following search parameters:
\begin{itemize}
    \item Symbol: ``\textbackslash{}sigma \_ \{ 8 \}''
    \item Unit: Dimensionless
    \item Value range: $0.4 \leq x \leq 1.5$
\end{itemize}
For this query we find 410 samples, with a mean of 0.828 and median of 0.803. Requiring uncertainties, we have 235 samples, with a mean of 0.808 and median 0.802.
There is little consensus amongst the result set on a \ann{ParameterName} string for this quantity, which is unsurprising, given the high linguistic variability seen for this parameter's name.
A plot of the collected measurements is seen in Fig.~\ref{fig:sigma8}\footnote{The results of this query may be found at: \url{http://numericalatlas.cs.ucl.ac.uk/constant/sigma8}}.

Here we see a clear convergence over time to a value of $\sim0.8$, as expected from the current understanding on the value of $\sigma_8$, with seemingly minimal tension across the years. A slight downward trend in the value of $\sigma_8$ is observed since around 2005.
Additionally, there is a clear drop-off in the number of reported measurements over the years since 2010. This is possibly due to an uptake in the use of $S_8$ (given by $\sigma_8 ( \Omega_M / 0.3 )^{0.5}$) over $\sigma_8$ in the literature.

\subsubsection{Baryon Density Parameter, $\Omega_b h^2$}
\begin{figure}
    \centering
	\includegraphics[width=0.9\columnwidth]{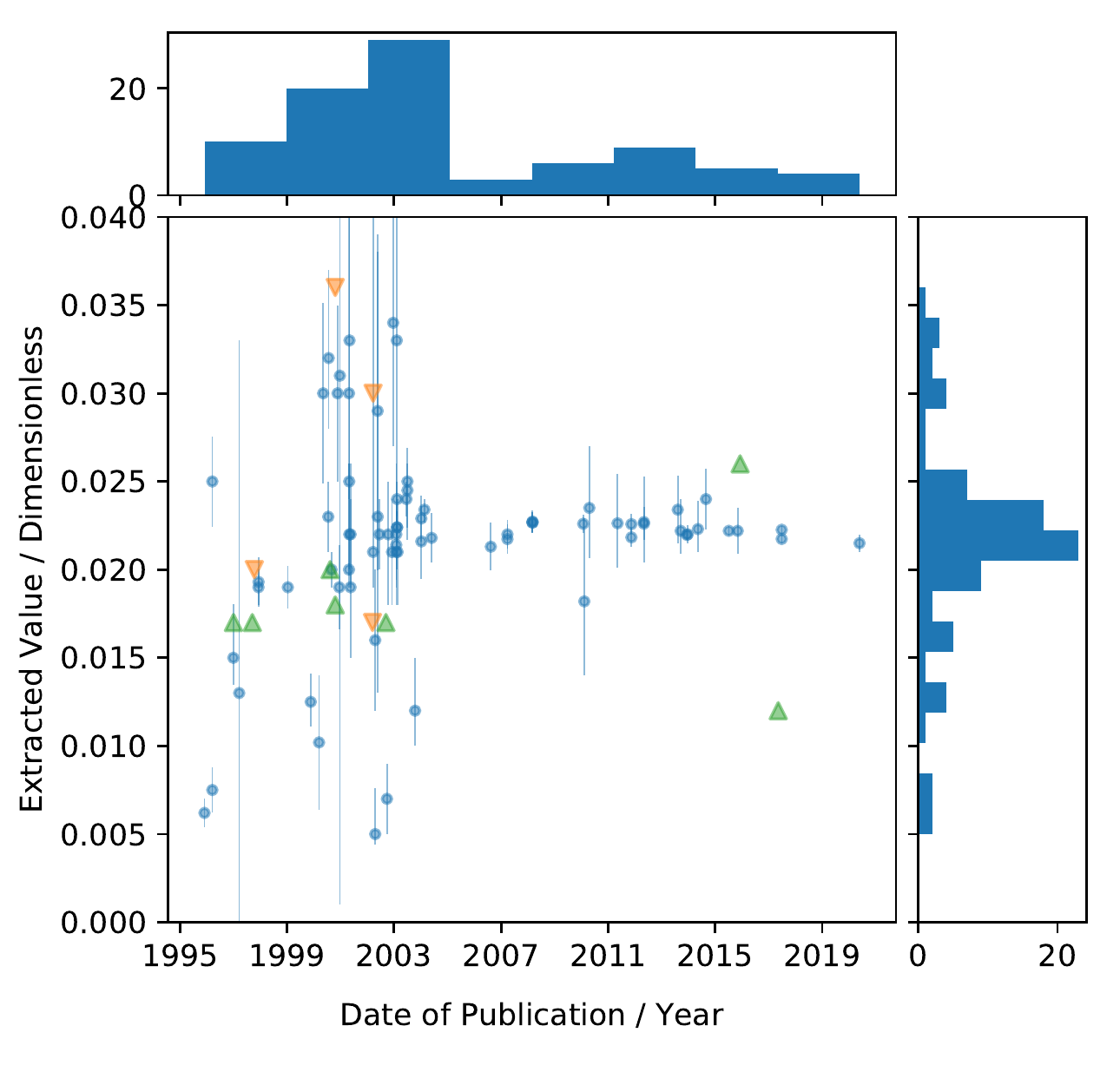}
	\caption{Time series of the search results for $\Omega_{b} h^2$, showing reported value and publication date.}\label{fig:omegabh2}
\end{figure}

In order to demonstrate the capacity of the model to recognising parameter symbols composed of multiple terms, we show the results for the baryon density parameter, $\Omega_b h^2$. The final search parameters are as follows\footnote{The results of this query may be found at: \url{http://numericalatlas.cs.ucl.ac.uk/constant/omegab}}:
\begin{itemize}
    \item Name: ``baryon density''
    \item Symbol: ``\textbackslash{}Omega \_ \{ B \} h \textasciicircum{} \{ 2 \}'', ``\textbackslash{}Omega \_ \{ b \} h \textasciicircum{} \{ 2 \}''
    \item Unit: Dimensionless
    \item Value range: $0.00 \leq x \leq 0.04$
\end{itemize}
Resulting in 86 measurements with provided uncertainties, with a mean of 0.0215 and a median of 0.022, as shown in Fig.~\ref{fig:omegabh2}. There is a clear consensus reached around 2003, possibly due to the WMAP publication in that year.
This result demonstrates that the model can identify compound symbols in the text (i.e. parameter symbols comprised of more than one syntactic component).

\subsubsection{Primordial Spectral Index, $n_{\rm s}$}
\begin{figure}
    \centering
	\includegraphics[width=0.9\columnwidth]{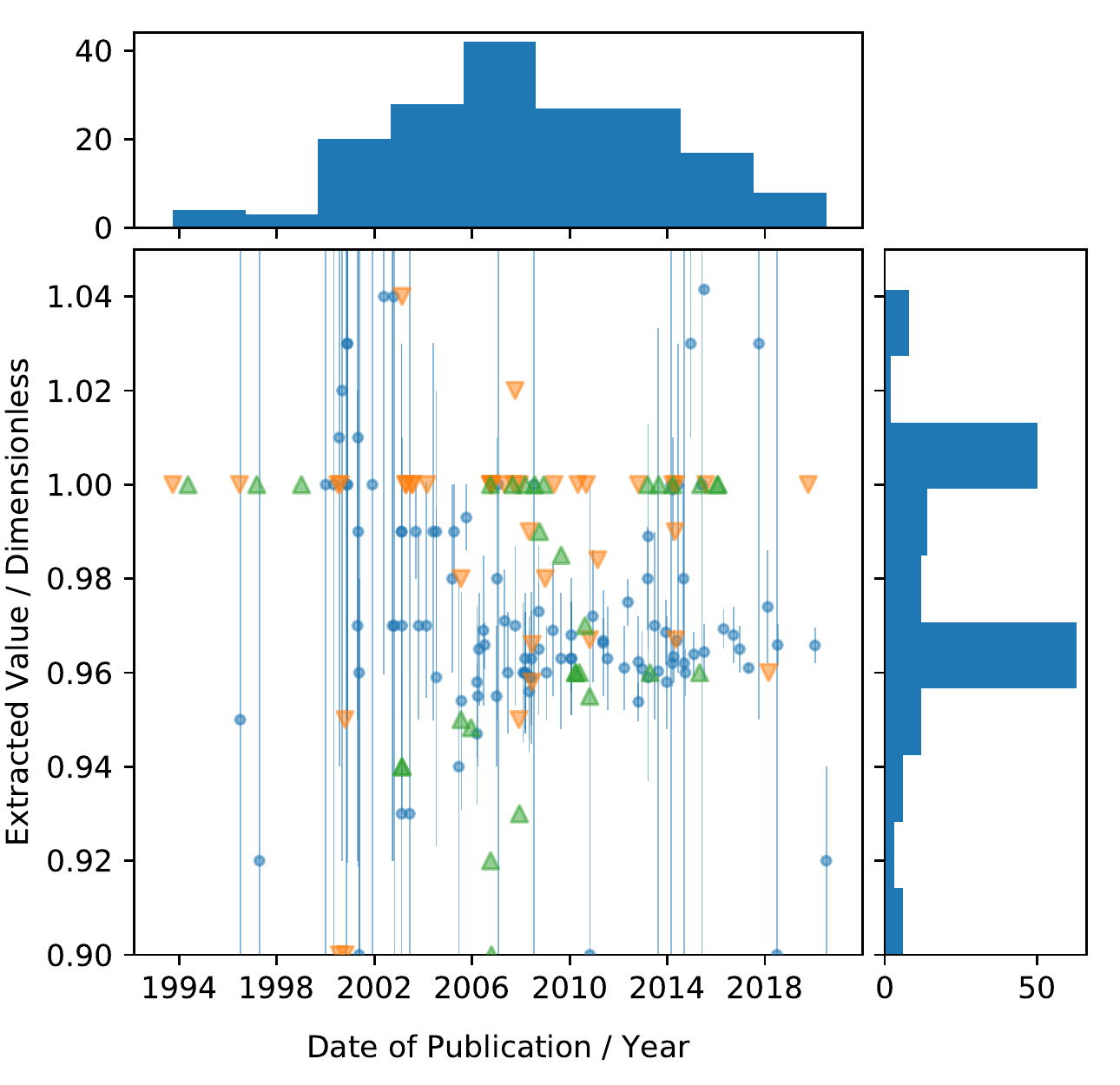}
	\caption{Time series of the search results for $n_{s}$, showing reported value and publication date.}\label{fig:spectralindex}
\end{figure}

For the primordial spectral index, $n$, the final search parameters are as follows\footnote{The results of this query may be found at: \url{http://numericalatlas.cs.ucl.ac.uk/constant/spectralindex}}:
\begin{itemize}
    \item Name: ``spectral index''
    \item Symbol: ``n \_ \{ s \}''
    \item Unit: Dimensionless
    \item Value range: $0.9 \leq x \leq 1.05$
\end{itemize}
Here experimentation was required to find a clean result set, as the symbol ``n'' (as is sometimes used for primordial spectral index) is far too common to be of use in discriminating the desired measurements from other parameters. Using a simpler name for the parameter also lead to a more productive search (as many instances in cosmology papers only state ``spectral index'', rather than ``primordial spectral index''). This search resulted in 100 measurements with provided uncertainties, with a mean of 0.972 and a median of 0.967. The plot for this result set is shown in Fig.~\ref{fig:spectralindex}.

A notable feature of this plot is the large number of constraint values at 1.0. Many of these are erroneous, or misleading -- for example, many are simply expressing very general statements about assumed cosmologies. However, if we examine the trend of central value measurements, we may note some interesting features: Firstly, we note that values with $n_s > 1$  are not seen after the start of 2003 (except a trio of values around 2015, which are incorrectly identified, and are in fact measurements of other physical quantities), coinciding with the WMAP 1 Year Data Release \citep{WMAP2003}. By the publication of the WMAP 3 Year Data Release \citep{WMAP2006} we see a much more cohesive set of results being reported (both in terms of value range and reported uncertainties), and the spread of values continues to narrow through to the present.
Whilst there appears to be a shift in uncertainty range during the 2013-16 period, many of these results are erroneous (``spectral index'' measurements relating to other physical quantities, generally), with the few correctly identified measurements either being discussions of different inflation models \citep{1308.4212, 1406.3243} or using some new technique for probing the cosmology \citep[e.g.][using cosmic voids]{1409.3364}.

\subsubsection{Sum of Neutrino Masses, $\sum m_\nu$}
\begin{figure}
    \centering
	\includegraphics[width=0.9\columnwidth]{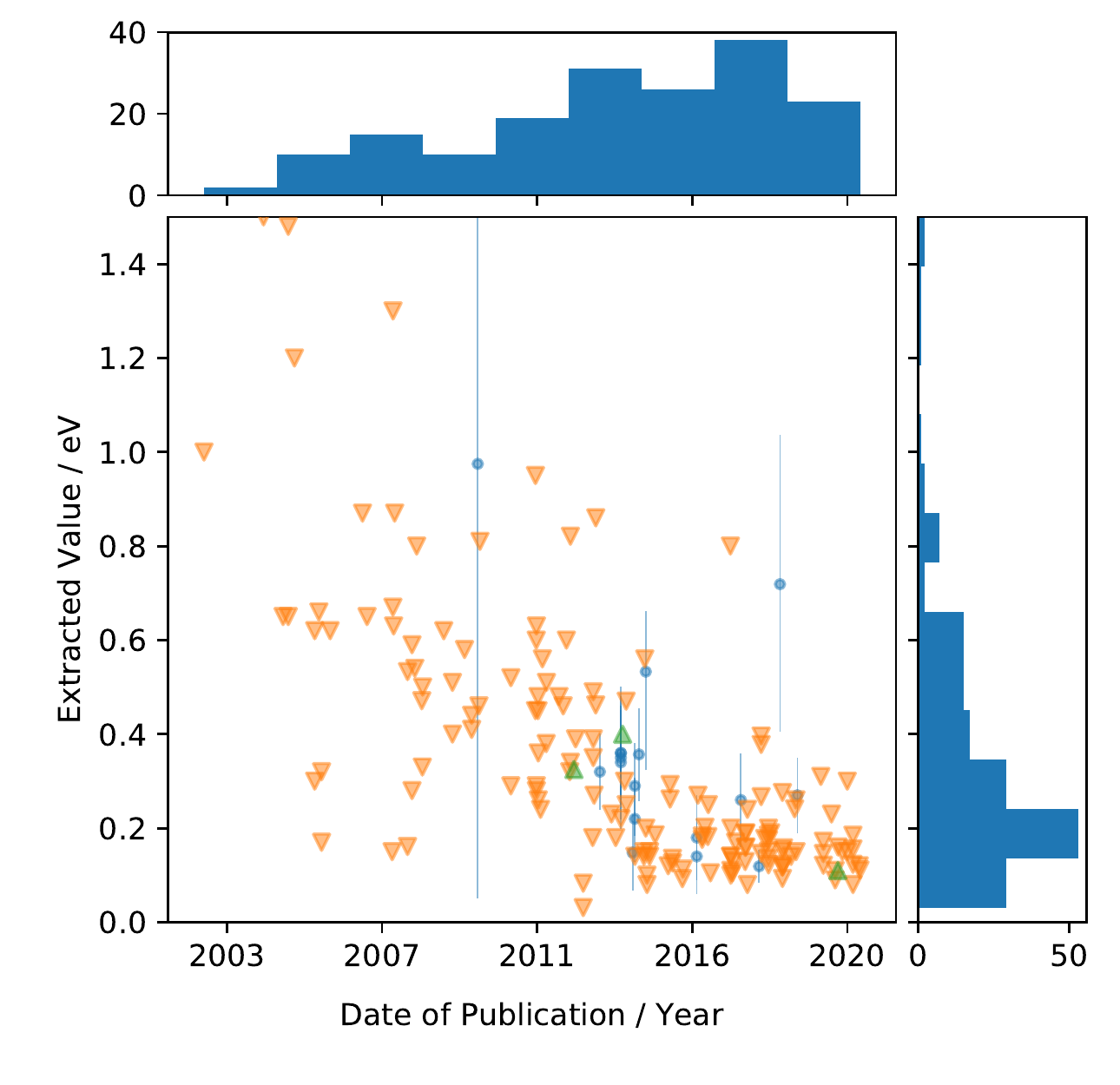}
	\caption{Time series of the search results for $\sum m_{\nu}$, showing reported value and publication date.}\label{fig:totalmnu}
\end{figure}

For the sum of neutrino masses, $\sum m_\nu$, the final search parameters are as follows\footnote{The results of this query may be found at: \url{http://numericalatlas.cs.ucl.ac.uk/constant/mnu}}:
\begin{itemize}
    \item Name: ``sum of neutrino masses'', ``total neutrino mass''
    \item Symbol: ``\textbackslash{}sum m \_ \{ \textbackslash{}nu \}'', ``\textbackslash{}sum M \_ \{ \textbackslash{}nu \}'', ``\textbackslash{}Sigma m \_ \{ \textbackslash{}nu \}'', ``\textbackslash{}Sigma M \_ \{ \textbackslash{}nu \}''
    \item Unit: eV
    \item Value range: $0 \leq x \leq 1.5$
\end{itemize}
These results are seen in Fig.~\ref{fig:totalmnu}. Here we see the utility of distinguishing \ann{MeasuredValue} and \ann{Constraint} annotations, as this is a quantity which is generally expressed as a constraint rather than a central value. However, it also presents another interesting challenge with regards to inferencing: there is an implied lower bound (i.e. zero) on the measurements which is not explicity stated. This is a natural assumption for a physicist reading the document, but one that relies on additional knowledge. As our future goals include automating aspects of the analysis phase as well as data collection, it is worth noting that these unspoken bounds must be taken into consideration.

We may also note from the plot the decided shift in the upper bound on $\sum m_\nu$ occuring at the start of 2015. This is, presumably, the influence of the publication of the Planck 2015 results \citep{Planck2015}, which reported a lower value than had been previously accepted. However, we may also see that a trend towards lower values had been in progress since approximately 2010.

\subsubsection{Dark Energy Equation of State Parameter, $w_0$}
\begin{figure}
    \centering
	\includegraphics[width=0.9\columnwidth]{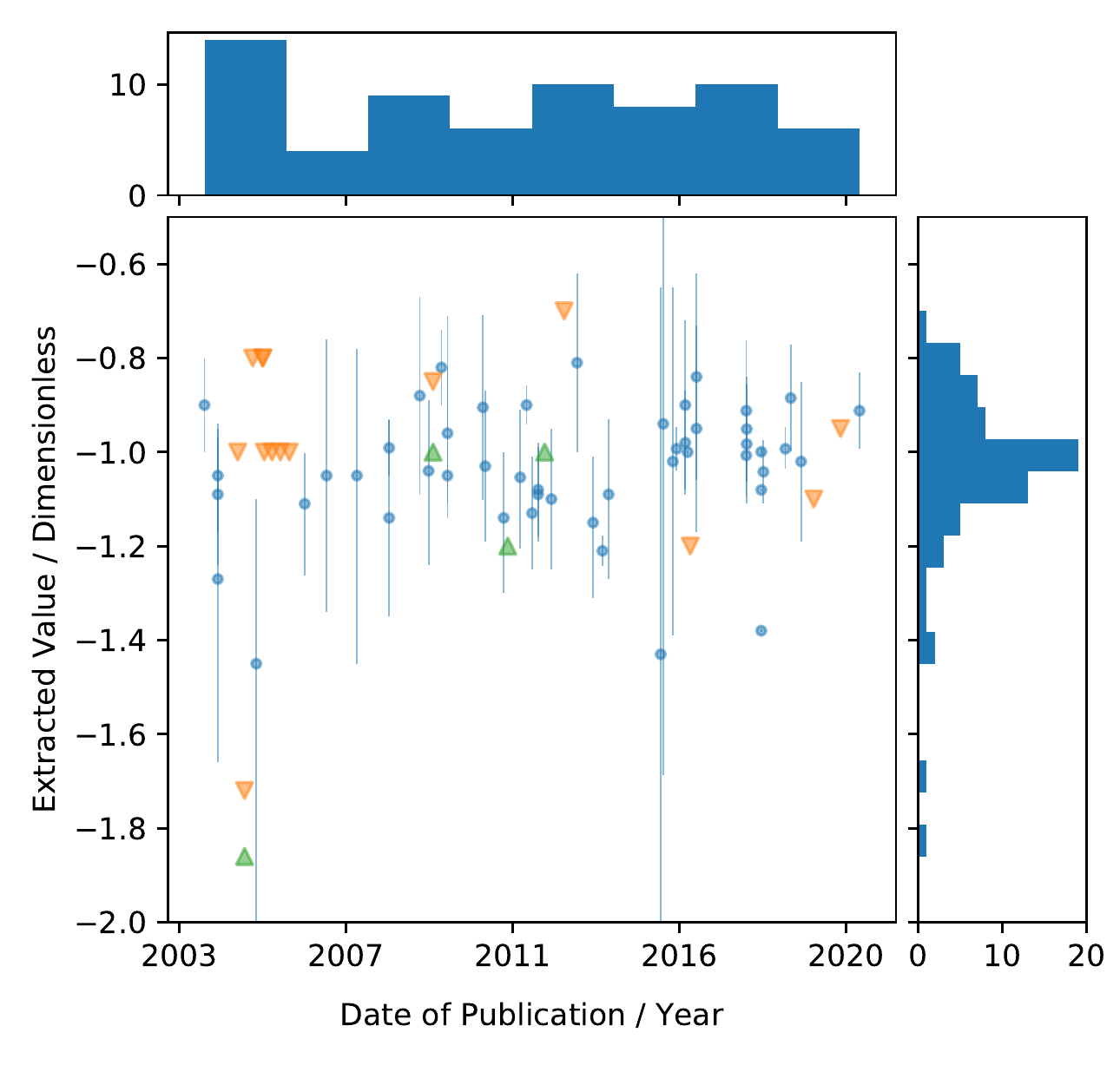}
	\caption{Time series of the search results for $w_{0}$, showing reported value and publication date.}\label{fig:w0}
\end{figure}

For the dark energy equation of state parameter, $w_0$, the final search parameters are as follows\footnote{The results of this query may be found at: \url{http://numericalatlas.cs.ucl.ac.uk/constant/darkenergyequationofstate}}:
\begin{itemize}
    \item Symbol: ``w \_ \{ 0 \}''
    \item Unit: Dimensionless
    \item Value range: $-2 \leq x \leq -0.5$
\end{itemize}
Resulting in 40 measurements with provided uncertainties, with a mean of -1.05 and a median of -1.05. Here we struggle with \ann{ParameterName} annotations, most likely due to a combination of the linguistic variability of this quantity's name, and the manner in which it is often reported (either simply as $w_0$, or cryptically as ``the equation of state parameter'' or similar). This makes it difficult to be certain that we have identified the correct values, beyond utilising some prior knowledge for the value range, considering the probability that the symbol ``w \_ \{ 0 \}'' may well be used in other contexts for different physical quantities. However, this being the case, the values collected by our search show a reasonable grouping, and the specialised nature of this parameter leads to a result set small enough to be easily examined manually. Plots of these results are shown in Fig.~\ref{fig:w0}.

There is a clear discontinuity in the plot after 2015, and examination of the papers following this shift suggest that this is due to new data from the Planck 2015 results \citep{Planck2015} and the SDSS Data Release 12 \citep{SDSSDR12} -- as can be seen in \citet{1601.03741, 1604.00183, 1604.06760, 1607.03151}. Additionally, there are several values reported over the years at approximately $-1.4$, which are found to be the result of investigations into different Dark Energy models \citep{astro-ph/0509082, 1802.05087} and cosmological measurements from GRBs \citep{1508.05898}.

\section{Conclusion}

We have presented our investigations into utilising artificial neural network models for extracting numerical astrophysics measurements from astrophysical literature. We have successfully trained neural models for Named Entity Recognition and Entity Attribute labelling tasks in this domain, and designed a rule-based approach for Relation classification based on the outputs of these neural models. The predictions from these models have been processed and structured to allow for searching based on a variety of criteria, such as parameter name or symbol, dimensionality, value range, and so on. During this process, we have created a hand-annotated training dataset for these tasks, based on paper abstracts from the arXiv repository.

We have compared the results from these new models to those of the model from our previous work \citep{crossland1}, and determined that there is significant overlap between the two result sets for our simple case study (the Hubble constant, $H_0$), showing that the new models have maintained the capabilities of the previous rule-based approach for simple cases. We then went on to show that the new models can be applied to a much broader range of scenarios, with a variety of different complexities, such as: dimensionless quantities, symbols which commonly occur in compound expressions (such as $\Omega_{m}$ occurring in $\Omega_{m} h^{2}$), or quantities with complex linguistic names (c.f. $\sigma_{8}$). We have shown that in these cases, with only a small amount of prior knowledge being leveraged in the search, a useful result set can be obtained, providing an excellent basis for further manual investigation or statistical analysis.
The database framework ensures very fast access to the model outputs, with each of the example queries requiring only seconds of compute time, allowing for quick iterations of search parameters in order to arrive at the desired result set.

Our results have been made available via an online interface, allowing users to search for parameters of interest with a variety of search criteria. Users will be able to engage with search results in an interactive manner, and download full result sets for their own experimentation and analysis. This interface, \textit{Numerical Atlas}, can be found at \url{http://numericalatlas.cs.ucl.ac.uk}. 
However, the numerical data are only one aspect of the model results. With the possibility of combining additional data from paper citations and references (e.g. from arXiv or NASA ADS), examining common naming conventions for symbols for use in other search environments, or finding common dimensions for a given parameter, there are many possibilities for examining the sociology and practices of the astrophysics community with this data.

With the extension of the capabilities of our model come some additional complexities. Firstly, there is still a large amount of noise present in the results, due to the intrinsic complexities of dealing with text. As we are now using neural models, these failure states appear less predictable to a human observer, in comparison to the output of rule-based models. Refining these models, and the pre- and post-processing steps used in our pipeline, will be an on-going task, involving the collection of additional training data and exploration of other potential model architectures and pipelines.


However, a bigger challenge than failure cases is dealing with the large variation seen in successfully extracted measurements - especially where parameter names and symbols are concerned. Our current strategy has involved extracting ``parameter names'' as single atomic entities. However, this is not a complete representation of the ``name''. For example, ``Galactic radius'' and ``radius of the Galaxy'' are, to an astronomer, clearly referencing the same physical quantity. However, this kind of entity normalization is a non-trivial task for machines. Currently we are relying on simple inclusion-based string matching, but this has many drawbacks -- in the above case, the only word shared between both forms (``radius'') is far too common to be sufficiently discriminative for a large scale search. The ability to automatically determine if two written names reference the same physical quantity (referred to as Entity Linking in the field of Natural Language Processing) would be a great boost to the practical utility of our search tool. More than this, such an analysis of the textual names would lead to more refined information on the nature of the parameters, as many names in scientific literature are grammatically descriptive (not all, of course -- there are plenty of ``Proper Noun constants'' to be found). For example, a grammatical breakdown of a name such as ``star-formation rate'' provides additional insight into the nature of the quantity: it is a \emph{rate} of some kind, relating to \emph{stars} and their \emph{formation}. With this breakdown, we could now search for parameters relating to stellar phenomena, and ``star-formation rate'' would be included in our listing. Naturally, this is a simplistic case, but the ability to search for parameters at a more abstract level would have many benefits.

Beyond additional processing of information we are currently collecting, there is also still much scope for collecting additional contingent information. The most important, perhaps, is the collection of experimental methodology. This task is complicated by the fact that it is generally a summarisation task -- where a ``Methodology'' section must be read and condensed down into a more compact description (ideally comprehensible to a human as well as the machine). In many cases there is no discrete method name provided at all (by the text itself, or indeed the community), and it is also possible that a paper is reporting a unique or ground-breaking experimental technique for which no term has yet been coined. There are certain sub-domains where a finite set of experimental techniques is available and well documented, but this is not the general case, and hence a more general approach must be found.

\section*{Acknowledgements}

This work was supported by the Science \& Technology Facilities Council (STFC Grant ST/N000811/1, STFC Doctoral Training Grant ST/R505171/1). CP acknowledges the financial support from the UCL Cosmoparticle Initiative.





\bibliographystyle{mnras}
\bibliography{ms} 




\appendix

\section{Detailed Annotation Schema Description}\label{app:anndescriptions}

Here we present an exhaustive list of descriptions of the annotation types used for the annotation effort described in Section~\ref{sec:annotationproject}.

For the Entity annotations, we have the following (as summarised in Table~\ref{tab:entities}):

\begin{itemize}
    \item \ann{MeasuredValue}: This Entity is used for the value, uncertainty, and units of numerical measurements reported as a central value with or without an accompanying uncertainty, when they appear together as a contiguous span in the text (e.g. ``5'' or ``5 \textbackslash{}pm 2''). This includes any textual notes which may appear inside the measurement, (e.g. ``5 \textbackslash{}pm 2 ( random ) km''), but does not include confidence limits -- unless they are stated within the bounds of the measurement (e.g. ``5 \textbackslash{}pm 2 ( 68 \% C.L. ) km'').
    \item \ann{Constraint}: This Entity is used for the value and units (and occasionally uncertainty) of constraints (such as the span ``0.42'' in ``\textbackslash{}alpha < 0.42''), where they appear as a contiguous span, not including any equality signs which may be present -- the nature of the constraint is provided by the \ann{UpperBound} and \ann{LowerBound} Attributes (discussed below). \textit{Note:} without the accompanying context, these often resemble instances of \ann{MeasuredValue}.
    \item \ann{ParameterName}: This Entity is for the linguistic name (i.e. a name comprised primarily of words, rather than symbols) of a measureable quantity. A measurement of the quantity does not have to be provided in the text for this annotation to be present. The exact span for such an Entity can be ambiguous, and can overlap with \ann{ObjectName} (see below). Parameter names can also be phrases, rather than simple nouns (or collections of nouns), and discussion between annotators is sometimes required to determine the exact start and end points of these Entities.
    \item \ann{ParameterSymbol}: This Entity is for the mathematical symbol for a physical quantity. These symbols can sometimes include abbreviations or short text strings (e.g. ``M \_ \{ vir \}''), but should not include complete words or phrases. They may also include associated brackets and their contents (e.g. ``H ( z = 0.36 )''), or less strictly mathematical syntax which is nonetheless a symbolic form (e.g. ``[Fe/H]''). Compound symbols (e.g. ``\verb|\Omega_m h^2|'') are accepted in cases where they are used as the primary identifier for a quantity, but compound mathematical expressions (or equations) which do not directly refer to a measured value should be annotated as \ann{Definition} (see below).
    \item \ann{ObjectName}: This Entity is used for the names (e.g. ``Milky Way'') or identifiers (e.g. ``M31'') of physical objects -- usually stars, galaxies, planets, etc. In some cases, if a sentence is structured such that a less concrete single object is discussed as one (for example, measurements of neutrino masses), then these physical entities may also be annotated as \ann{ObjectName}.
    \item \ann{ConfidenceLimit}: This Entity is for the numerical value and quantifier (usually ``\textbackslash{}sigma'' or ``\%'') of confidence limits (excluding any accompanying phrase, such as ``C.L.''). This Entity is required due to the fact that many reports of confidence limits are separated by some span from the measurement they refer to - and often a single instance of a confidence limit is used for multiple measurements.
    \item \ann{SeparatedUncertainty}: This Entity is used for an uncertainty provided separately from its central value. This annotation is required as we discovered several instances in which a value and its corresponding uncertainty occur at different points in the text -- this is usually where the calculation of the measurement uncertainty was non-trivial in and of itself.
    \item \ann{Definition}: This Entity is for mathematical expressions or equations which are comprised of more than one symbol, and which are stated in the text as formulae, rather than contained mathematical statements.
\end{itemize}

For the Relation annotations (as summarised in Table~\ref{tab:relations}):

\begin{itemize}
    \item \ann{Measurement}: This Relation indicates that a \ann{MeasuredValue} or \ann{Constraint} is a direct numerical measurement of some stated parameter (\ann{ParameterName} or \ann{ParameterSymbol}). This Relation should only be used for direct instances of the value of the parameter in question, not derived quantities or contingent values.
    \item \ann{Name}: This Relation is used to indicate that a \ann{ParameterSymbol} is a mathematical expression for a linguistic name (\ann{ParameterName}) found in the text.
    \item \ann{Confidence}: This Relation indicates that a \ann{ConfidenceLimit} annotation is related to a measurement annotation -- i.e. that the stated confidence limit relates to the uncertainties provided in the measurement. This Relation should only be used for \ann{MeasuredValue} annotations which provide an uncertainty, but can be used for any \ann{Constraint} annotation.
    \item \ann{Property}: This Relation indicates that a measurement (\ann{MeasuredValue} or \ann{Constraint}) or parameter (\ann{ParameterName} or \ann{ParameterSymbol}) is a direct property of an object specified by an \ann{ObjectName} annotation. This generally means that the parameter is a physical characteristic of the object (``mass'', ``radius'', etc), or that it represents some important property associated with the object (e.g. ``star-formation rate'' of a galaxy).
    \item \ann{Equivalence}: This Relation indicates that two \ann{ObjectName} annotations (with different textual contents) relate to the same physical object.
    \item \ann{Contains}: This Relation indicates that one object \textit{contains} another object. This could be used for sub-components of a system (e.g. members of a binary star system), or objects which reside within a larger object (e.g. stars within a galaxy).
    \item \ann{Uncertainty}: This Relation exists to connect \ann{MeasuredValue} or \ann{Constraint} annotations to a \ann{SeparatedUncertainty} annotation, indicating that the uncertainty is directly related to the measurement. This should only be used where the value and uncertainty share the same dimensions, and require no additional manipulation to be used together.
    \item \ann{Defined}: This Relation indicates that a \ann{Definition} annotation contains a mathematical definition for another Entity. This is often of the form ``y = mx + c'', but could be more verbose (e.g. ``\textbackslash{}alpha, which is defined to be ...'').
\end{itemize}

And finally for the Attribute annotations (as summarised in Table~\ref{tab:attributes}):

\begin{itemize}
    \item \ann{Incorrect}: This Attribute is applied to measurement annotations which are stated to be incorrect by the author (regardless of whether the author's determination is true).
    \item \ann{AcceptedValue}: This Attribute indicates that a given measurement annotation is stated as final, or ultimately accepted, by the author -- as may occur in cases where several possible numerical values are provided based on different assumptions.
    \item \ann{FromLiterature}: This Attribute indicates that a measurement is not the work of the author, but instead quoted from some literature source.
    \item \ann{UpperBound}: This Attribute indicates that a \ann{Constraint} annotation represents an upper bound on a quantity.
    \item \ann{LowerBound}: This Attribute indicates that a \ann{Constraint} annotation represents a lower bound on a quantity.
\end{itemize}

\section{Consensus Annotation Algorithm}\label{app:consensusalgorithm}

For the collection of annotated paper abstracts to be used as training data for machine learning purposes, we must consolidate the repeated sets of annotations for each abstract (see Section~\ref{sec:annotationproject}) into a single annotation set for that particular piece of text. This should be done in such a way that we preserve the largest amount of information from the annotators, while also taking account of ambiguity and guarding against human error. There is not necessarily a canonical approach to take for this problem, and so we have chosen the following method:

For each abstract, $D$, with a set of annotations, $S$, consisting of Entities, $E$, Relations, $R$, and Attributes, $A$, we group the Entities into overlapping groups. Each of these groups can be in one of several states: full agreement, partial agreement, or disagreement. In the case of full agreement, all annotators have exactly the same Entity annotations (both the span of the annotation and it's label), and this annotation is accepted into the consensus annotation set. For partial agreement, more than half the annotators (2 in our case) must have the same annotation, and this is also considered a consensus annotation. For the disagreement case there are many possible situations: the annotators may all have different overlapping spans with the same label, selected different labels for the same span, multiple sets of partially overlapping spans, or some combination thereof. It is also possible that a single annotation span for one annotator may be multiple spans for another, or that only one annotator identified a certain span as containing an Entity, and other such combinations of labelling.
One of these cases is resolved by the consensus algorithm in the following way: If more than half the annotators have overlapping annotation spans with the same label, which do not intersect with any other spans (i.e. we are not in a case where one annotator has a single span and another multiple spans in the same region), then a consensus Entity is created from the overlap of the annotated spans, and assigned the appropriate label (these substitutions are tracked for the purposes of consensus Relations -- see below).
For all other cases, the annotation is simply rejected from the consensus.

Next we consider Relations: first we filter the candidate Relations by whether their start and end Entities are in the consensus set -- if not, the Relation is rejected. The remaining Relations are then grouped together by their start and end Entities, and the same process of identification as full agreement, partial agreement, or disagreement is performed. However, for Relations, the possible combinations of agreement and disagreement are less complex. A simple majority (2, in this case) voting system is sufficient to determine inclusion in the consensus set.

Finally, Attribute annotations are also filtered by subject Entity inclusion in the consensus set, and agreement is determined by voting, as for Relations.



\section{Rule-based Relation Model Details}\label{app:rulesreldetails}

For the direct text evaluation, we have the following rules:
\begin{itemize}
    \item Any \ann{ParameterSymbol} and \ann{MeasuredValue} separated exactly (ignoring whitespace) by one of: ``='', ``>'', ``<'', ``\textbackslash{}sim'', ``\textbackslash{}simeq'', ``\textbackslash{}approx'', ``\textbackslash{}leq'', ``\textbackslash{}geq'', ``of'', or an empty string (i.e. whitespace) are considered to be linked by a \ann{Measurement} Relation.
    \item Any \ann{ParameterName} and \ann{MeasuredValue} separated exactly by one of: ``is'', ``of'', ``('', or an empty string are considered to be linked by a \ann{Measurement} Relation.
    \item Any \ann{MeasuredValue} and \ann{ConfidenceLimit} separated exactly by one of: ``at'', ``at the'', or ``('' are considered to be linked by a \ann{Confidence} Relation.
    \item Any \ann{ParameterName} and \ann{ParameterSymbol} separated exactly by: ``is'', ``is the'', ``of'', ``('', a comma, or an empty string are considered to be linked by a \ann{Name} Relation.
    \item Any \ann{ParameterName} and \ann{ObjectName} separated exactly by: ``of'', ``of the'', or an empty string are considered to be linked by a \ann{Property} Relation.
    \item Any \ann{ParameterSymbol} (or \ann{ParameterName}) and \ann{Definition} separated exactly by: ``is'', ``='', or ``\textbackslash{}equiv'' are considered to be linked by a \ann{Defined} Relation.
\end{itemize}

Next, for the Entity patterns, we use the following rules (these patterns only consider the sequences of Entities in a sentence, and ignore all other tokens labelled as \ann{None})):
\begin{itemize}
    \item Simple Name pattern: A \ann{ParameterName} followed by a \ann{ParameterSymbol} (but not preceeded by one) is considered to be linked to it by a \ann{Name} Relation.
    \item Multiple Measurements pattern: A \ann{ParameterName} or \ann{ParameterSymbol} annotation followed by a series of \ann{MeasuredValue} or \ann{Constraint} annotations is considered to be linked to each by a \ann{Measurement} Relation.
    \item Standard Measurement pattern: A \ann{ParameterName} followed by a \ann{ParameterSymbol} followed by a \ann{MeasuredValue} or \ann{Constraint} annotation are assumed to be linked by \ann{Name} and \ann{Measurement} Relations.
    \item Definition Measurement pattern: A \ann{ParameterSymbol} followed by a \ann{Definition} followed by a \ann{MeasuredValue} is considered to have the \ann{ParameterSymbol} linked to the other two by a \ann{Defined} and a \ann{Measurement} Relation, respectively.
    \item Simple Confidence Limit pattern: Any series of \ann{MeasuredValue} or \ann{Constraint} annotations followed by a single \ann{ConfidenceLimit} are each considered to be linked to it by a \ann{Confidence} Relation.
    \item Named Object Property pattern:  A \ann{ParameterName} followed by an \ann{ObjectName} followed by a \ann{ParameterSymbol} followed by a \ann{MeasuredValue} or \ann{Constraint} annotation are considered to be linked by a \ann{Property} Relation between the \ann{ObjectName} and \ann{ParameterName}, and by a \ann{Name} Relation between the \ann{ParameterName} and \ann{ParameterSymbol} (the \ann{Measurement} Relation is already covered by above rules).
    \item Simple Property pattern: An \ann{ObjectName} followed by a \ann{ParameterName} and/or \ann{ParameterSymbol} annotation, followed by a \ann{MeasuredValue} or \ann{Constraint} annotation, are considered to be linked by a \ann{Property} Relation, optionally a \ann{Name} Relation (if both name and symbol are present), and finally by a \ann{Measurement} Relation.
    \item Tuple Measurements pattern: An uninterrupted sequence of \ann{ParameterName} and \ann{ParameterSymbol} annotations, followed by another uninterrupted sequence of \ann{MeasuredValue} and \ann{Constraint} annotations, of equal length, are considered to be pairwise linked by \ann{Measurement} Relations. This pattern is commonly seen when reporting multiple values from cosmological simulations (often collections of cosmological parameters).
\end{itemize}

Using the above rules, a reasonable degree of accuracy can be achieved on the annotated data available, as may be seen in the results section below.

\section{Annotation Post-Processing Steps}\label{app:postprocessing}

The following steps are carried out for the Entity annotations in the prediction documents (note that, in practice, these post-processing steps are performed before Attribute or Relation predictions are performed):
\begin{enumerate}
    \item As for the consensus algorithm (see Appendix \ref{app:consensusalgorithm}), stopwords are removed from the beginning and end of all Entities. Unlike the consensus dataset, here we may run into cases where this removes the entire Entity string (as false positives containing only stopwords -- e.g. ``of'', or `'of the'' -- are a standard error encountered in the predictions), and in these cases the Entity annotation is completely discarded.
    \item Any \ann{ParameterSymbol}, \ann{ParameterName}, or \ann{ObjectName} annotations which do not include at least one alphabetical character are discarded.
    \item Any \ann{MeasuredValue} or \ann{Constraint} annotations which do not include at least one numerical character are discarded.
    \item Again, as for the consensus algorithm, any repetitions of the textual content of any \ann{ParameterSymbol}, \ann{ParameterName}, or \ann{ObjectName} annotations are identified in the document and annotated with the corresponding Entity label.
\end{enumerate}

We also use parsing algorithms to normalise certain Entities -- notably \ann{ConfidenceLimit} and \ann{ParameterSymbol} Entities. For \ann{ConfidenceLimit} annotations we perform a simple pattern match with a regular expression, requiring the text to follow one of the following patterns (with allowances for some minor variations of whitespace, etc.):
\begin{itemize}
    \item $1 \sigma$
    \item 1-$\sigma$
    \item one sigma
    \item one-sigma
    \item 68\%
    \item 68 percent
\end{itemize}
Percentage expressions of the confidence are converted into standard deviations using the inverse error function. This information will allow for measurement errors to be converted into a standard format by the user if desired.

The normalisation process for \ann{ParameterSymbol} annotations is a little more complex, due to the repeating and recursive nature of \LaTeX{} symbols and mathematical expressions in general. The goal is to normalise the string representation of the symbol such that different typographical forms of the same mathematical symbol can be better compared using standard string comparison. For example, we desire that the following strings be considered equal for our search:
\begin{itemize}
    \item ``H \_ 0'' and ``H \_ \{ 0 \}'' (as \LaTeX{} does not require braces for single character sub- and super-scripts)
    \item ``Fe/H'' and ``Fe / H'' (whitespace differences like this can occur when symbols can be written in \texttt{math} or \texttt{text} environments, causing the symbol to be parsed into a different number of tokens)
    \item ``T \_ \{ \textbackslash{}mathrm \{ eff \} \}'' and ``T \_ \{ eff \}'' (the \LaTeX{} command here is aesthetic, and does not indicate a different semantic meaning)
    \item ``a / b'' and ``\textbackslash{}frac \{ a \} \{ b \}''
    \item ``f ( x )'' and ``f \textbackslash{}left( x \textbackslash{}right)''
\end{itemize}
Note that the increased quantity of whitespace characters in these examples simply reflects the tokenizing of the source \LaTeX{} by our pre-processing pipeline.

In order to normalize these highly variable strings we have created a context free grammar with which to parse the raw \LaTeX{} strings into a recursive tree structure representing the components of the symbol -- for example, individual characters, sub- and superscript symbols, functions (i.e. ``$f\left(x\right)$''), bracketed expressions (respecting bracket type), binary operator expressions (e.g. ``a + b''), and so on. These data structures may then be serialized into a standard string format, obeying \LaTeX{} style conventions. There are many cases where this parsing fails, either because the symbol represents some typographic edge case, or because the span identified by the model is incomplete. Currently no attempt is made to alter the span of the Entity in question, and failed parsing attempts result in the original string also being used to represent the normalised case for search purposes. This normalised string may then be used for queries based on mathematical symbols, with the query symbol string also being passed through this parsing algorithm to ensure that it is inline with the expected style conventions -- for example, braces (``\{\}'') are included in all cases of ambiguity (i.e. ``H \_ 0'' becomes ``H \_ \{ 0 \}''), and mathematical (as opposed to \LaTeX{} type-setting) braces use their simplest form (i.e. ``\textbackslash{}right('' becomes ``('') to improve readability for the user.

For the Relation annotations, as with the consensus dataset (see Appendix \ref{app:consensusalgorithm}), we add any transitive or implied Relations into the annotation set for the document. This is especially crucial at this stage, as having all implied Relations be present in the annotations makes search-time operations more efficient, by removing the need for further inferencing at a later stage.


\bsp	
\label{lastpage}
\end{document}